\documentstyle[a4,psfig,epsf]{article}

\newcommand \beq{\begin{eqnarray}}
\newcommand \enq{\end{eqnarray}}
\newcommand \be{\begin{equation}}
\newcommand \en{\end{equation}}

\newcommand{\mafigura}[4]{
  \begin{figure}[hbtp]
    \begin{center}
      \epsfxsize=#1 \leavevmode \epsffile{#2}
    \end{center}
    \caption{#3}
    \label{#4}
  \end{figure} }   

\begin{document}
\title{Nucleon-nucleon effective potential in dense matter including 
rho-meson exchange.}
\author{L. Mornas$^{(a)}$, E. Gallego$^{(b)}$ and A. P\'{e}rez$^{(b)}$ \\
{\small{\it $(a)$ Departamento de F{\'\i}sica, Universidad de Oviedo, E-33007
Oviedo (Asturias) Spain}} \\
{\small{\it $(b)$ Departamento de F\'{\i}sica Te\'{o}rica Universidad 
de Valencia, E-46100 Burjassot (Valencia) Spain. }} }
\maketitle

\begin{abstract}
We obtain the RPA summed one-meson exchange potential between nucleons in
symmetric nuclear matter at zero temperature, from a model which includes 
$\rho $, $\sigma $, $\omega $ and $\pi $ mesons. The behavior of rho mesons
inside the medium is first discussed using different schemes to extract
a finite contribution from the vacuum polarization. These schemes give
qualitatively different results for the in-medium rho mass. The results are
discussed in connection with the non-renormalizability of the model. We next
study the modified potential as density increases. In the intermediate
distance range, it is qualitatively modified by matter and vacuum effects.
In the long-distance range ($r>2$ fm), one observes the presence of
oscillations, which are not present in free-space. Features on this distance
range are insensitive to the renormalization scheme.
\end{abstract}



\section{Introduction}

Quantum Hadrodynamics (QHD) designates a class of models in which nuclear
interactions are described through {\it effective} relativistic Lagrangians
of nucleons coupled to different kinds of mesons. Such models have been
successfully used during the past decades to study very different
situations, such as nucleon-nucleon scattering processes in free space 
\cite{M89} or the nuclear many-body problem \cite{[SW86]} (for a recent 
review of QHD models, see \cite{[Se97]}). In vacuum, the one-boson exchange
approximation gives a reasonable approximation to nucleon-nucleon scattering
data. 

The situation becomes more complicated in dense nuclear matter
since higher-order diagrams have to be taken into account. It is known
for example that short range correlations arise due to the hard
repulsive core in the free potential. An important achievement
of the late eighties was the reproduction of saturation properties
by parameter-free, (relativistic)  Dirac-Brueckner G-matrix 
calculations \cite{[BM90],[BM96]}. In this approach, repeated 
exchange of free mesons give rise to an effective potential with
a smoothened repulsive core. On the other hand, it is also 
well-known that effects coming from the Random Phase Approximation 
(RPA) will modify the two-particle interaction at high densities. 

For example, in QED the lowest-order interaction potential between two 
static charges is the usual Coulomb potential. Inside a plasma, RPA 
effects modify the photon propagator, and correspondingly one obtains 
a screened Debye potential. At larger distances, new phenomena appear. 
If the temperature is sufficiently low, the potential becomes 
oscillatory and damped as some power-law of the distance. These are 
the so-called {\it Friedel oscillations} \cite{FR52}, and are 
originated by the sharp profile of the Fermi surface at low 
temperatures. Friedel oscillations are supported by a large body 
of experimental evidence in metallic alloys (see {\it e.g.} 
\cite{RO60}). 

Similar screening effects are encountered in a QCD quark-gluon plasma 
when the quark-antiquark potential is calculated within the same 
approximations \cite{KA88}. Such effects appear also if one considers 
the spatial dependence of static meson correlation functions at finite 
baryon density in the Nambu -- Jona -- Lasinio model \cite{FF96}.
In a nuclear plasma, the presence of the medium will also give rise 
to a modified interaction. The modifications of the potential due 
to RPA have been investigated so far in the case of the one-pion 
exchange \cite{DPS89} and for the Walecka $\sigma$ - $\omega$ model 
\cite{GDP94,GDP94L}. In both cases, a screened  potential was obtained 
that differs from the one in vacuum . At long distances 
($ r\geq 2$ fm), the potential shows also an oscillatory behavior, 
as in the QED and QCD cases. 

Our claim is that RPA may give important effects, both quantitative 
and qualitative, and has to be considered in many-body calculations 
of nuclear matter. As a matter of fact, it was shown \cite{DM98} 
that RPA corrections give rise to a modification of the in-medium 
cross section of the same order of magnitude as Brueckner ones, 
and compatible with experimental data available so far. Other examples 
are the reduction of neutrino opacities in neutron star matter from 
RPA corrections of the NN interaction \cite{IP82,RPLP99} and the 
measurement of the electromagnetic response function in quasielastic 
electron scattering experiments \cite{W93}.

Of course, both kinds of effects should eventually be considered. Within 
a diagrammatic perturbation expansion, they appear as summations of two
distinct subsets of diagrams, namely particle-hole loops for the RPA,
and ladder summation for the G-matrix mentioned above. Both are essential 
ingredients, since RPA summation is needed in order to describe long-range 
correlations, and, on the other hand, ladder summations are needed to 
create short-range correlations. 

It would be desirable to develop some approach to incorporate both kind of
summations in a systematic way. As discussed in \cite{JW94}, however, 
this can not be done simply by using the RPA modified potential as a 
driving term for ladder summation. Such an approach would lead to 
inconsistencies and double counting, and one has to look for more 
elaborated techniques. Parquet resummation \cite{JW94} is the minimal 
extension that can accomplish this purpose, but  its extension to
 relativistic strongly interacting systems still encounters 
technical difficulties. Besides theoretical problems, parquet techniques 
applied to interactions with a simpler structure, (scalar $\lambda
\phi^4$ or QED) already lead to very complex results exceeding standard 
computational resources and tricky convergence issues. 
\footnote{Interesting attempts in this direction may also be found in 
a non relativistic formalism in \cite{DS90,BKS99,HCP95}.}

Before embarking on such an attempt, our aim will be more modest.
We think that it is useful to check whether the RPA effects give rise to 
sizeable corrections in the density and momentum transfer ranges of 
interest. Actually, we already know from other contexts that the answer 
is positive \cite{DM98,RPLP99,CL94}. Here we focus on the analysis of 
the modifications of the one-meson exchange potential due to RPA 
corrections to the meson propagator inside nuclear matter. We will 
pay special attention to the comparison of the modified potential, 
in contrast to the free-space potential. Keeping in mind the above 
considerations, since the RPA summation is not the only effect on 
the many-body problem, it cannot be used in naive ladder calculations. 
We insist, however, that this summation is a necessary (and important) 
piece of the problem, and deserves further study. 
In the course of the calculations, we obtain the modified meson 
propagator inside the medium, a result which is interesting by itself, 
since it allows us to study the effect of the medium on the meson 
propagation, the modification of the meson mass, the appearance of 
new branches, {\it etc.}

In this paper, we concentrate on these modifications within a model which
includes $\rho $ mesons, in addition to the $\pi$, $\sigma $ and $\omega $
mesons already considered in the previous references. The inclusion of 
$\rho$ mesons is a necessary ingredient in determining the proper isospin
dependence of the nucleon-nucleon potential within meson exchange models 
\cite{M89}. Also, obtaining the rho propagator in dense matter will allow 
us to investigate how its in-medium mass changes as density increases. 
This issue has become important in the context of dilepton excess 
at low invariant masses observed in heavy ion collisions \cite{ChR96}. 
As we will show, however, the evolution of the $\rho $ meson mass with 
density in this QHD model is very sensitive to the way in which the 
vacuum terms of the polarization are treated. 
This is due to the non-renormalizability of the derivative part of the 
$\rho$ meson coupling which was introduced for phenomenological reasons.
While renormalizability might not be a requirement in the case of 
{\it effective} models, one has to provide a prescription to eliminate 
the divergences that arise in these relativistic models {\it consistently}.
As a matter of fact, one finds that, in renormalizable QHD models, 
vacuum effects eliminate the pathologies which appear when all vacuum 
terms are simply left out \cite{DP91}. We will examine here several 
different prescriptions, leading to different consequences for 
the rho mass in nuclear matter.

This paper is organized as follows. In section 2 we define the model and
calculate the corresponding RPA meson propagators in the medium. In section
3 we discuss how to extract a finite vacuum term from the vacuum
polarization in the case of rho mesons. We give in section 4 general
expressions for the one-boson exchange potential obtained from the modified
propagator. In section 5 we discuss the rho-meson dispersion relations 
and effective mass in connection with the renormalization
procedures discussed in section 3. Results concerning the in-medium
potential are showed in section 6. We first consider only the exchange of
rho mesons. After this, we give some plots which are obtained by adding all
kind of mesons present in our model. Mechanisms which could be responsible
for smearing away the Friedel oscillations present in this potential are 
discussed in Section 7. Finally, our main results are summarized and 
commented in Section 8. We have gathered in Appendix A and B the explicit 
formulae concerning the matter and vacuum part of the polarization tensor 
of rho mesons respectively. Appendix C contains the configuration space 
contributions from different mesons to the nucleon-nucleon potential.

\section{Meson propagators in symmetric nuclear matter.}

In this section, we describe the main steps necessary to obtain the meson
propagators in the RPA approximation. The starting point will be a QHD
model. For this, we adopt the one described by the following Lagrangian : 
\begin{equation}
{\cal L}\ =\ {\cal L}_{N}\ +\ {\cal L}_{\sigma }\ +\ {\cal L}_{\omega }\ +\ 
{\cal L}_{\pi }\ +\ {\cal L}_{\rho }\ +\ {\cal L}_{I}\ +\ {\cal L}_{CC}
\label{L}
\end{equation}
which assumes nucleons interacting with several kinds of mesons : $\sigma
,\pi ,\rho $ and $\omega $. In the latter equation, ${\cal L}_{N}$
corresponds to the nucleon Dirac free Lagrangian: 
\begin{equation}
{\cal L}_{N}\ =\ (i/2)\,[\,\ {\overline{\psi }}\gamma \cdot \partial \ {\psi 
}\,-\,(\partial \ {\overline{\psi }})\cdot \gamma \ {\psi }\,]\,-\,m\ {%
\overline{\psi }}\ {\psi }
\end{equation}
where $\psi =\left( 
\matrix{
\psi _{p} \cr
\psi _{n} \cr}
\right) $ is the nucleon isospin-doublet (protons and neutrons) field with
mass $m$, 
\begin{equation}
{\cal L}_{\sigma }\ =\ (1/2)\,[\,(\partial ^{\,\nu }\ {\sigma })(\partial
_{\,\nu }\ {\sigma })\,-\,\mu _{\sigma }^{\ 2}\ {\sigma }^{2}\,]
\end{equation}
\begin{equation}
{\cal L}_{\pi }\ =\ (1/2)\,[\,(\partial ^{\,\nu }\ {\vec{\pi}})(\partial
_{\,\nu }\ {\vec{\pi}})\,-\,\mu _{\pi }^{\ 2}\ {\vec{\pi}}^{2}\,]
\end{equation}
\begin{equation}
{\cal L}_{\rho }\ =\ -(1/2)\,[\,(1/2)\,\ {\vec{R}}_{\rho }^{\,\mu \nu
}\,\cdot \,{\vec{R}}_{\,\rho \mu \nu }\,-\,\mu _{\rho }^{\ 2}\,\ {\vec{\rho}}%
^{\,\nu }\,\cdot \,{\vec{\rho}}_{\nu }\,]\ 
\end{equation}
\begin{equation}
{\cal L}_{\omega }\ =\ -(1/2)\,[\,(1/2)\,\ {F}_{\omega }^{\,\mu \nu }\,\cdot
\,{F}_{\,\omega \mu \nu }\,-\,\mu _{\omega }^{\ 2}\,\ {\omega }^{\,\nu
}\,\cdot \,{\omega }_{\nu }\,]\ 
\end{equation}
are the free Lagrangians for the mesons, with masses $\mu _{i}$ (i=$\sigma
,\pi ,\rho $ and $\omega $). ${\cal L}_{I}$ gives the meson-nucleon
interaction. We have adopted simple Yukawa couplings for the $\sigma ,\pi $
and $\omega $ mesons. For the $\rho $-meson we added a tensor term \cite{M89}%
. Therefore, we have : 
\begin{eqnarray}
{\cal L}_{I}\ =\  &&g_{\sigma }\,\ {\overline{\psi }}\,\ {\sigma }\,\ {\psi }%
\,+\,g_{\omega }\,\ {\overline{\psi }}\,\gamma ^{\mu }\,\ {\omega }_{\mu
}\,\ {\psi }\ -\,i\ g_{\pi }\ \ {\overline{\psi }}\ \gamma ^{5}\ \vec{\tau}%
\cdot \ {\vec{\pi}}\ {\psi }  \nonumber \\
&&+\ g_{\rho }\ \ {\overline{\psi }}\ \gamma ^{\mu }\ \vec{\tau}\cdot \ {%
\vec{\rho _{\mu }}}\ {\psi }\ -\ {\frac{{f_{\rho }}}{{2m}}}\ \ {\overline{%
\psi }}\ \sigma ^{\mu \nu }\ \ {\psi }\ \vec{\tau}\cdot \partial _{\nu }\ {%
\vec{\rho _{\mu }}}  \label{LI}
\end{eqnarray}
with the following notations : 
\begin{equation}
\ {F}_{\,\omega }^{\,\mu \,\nu }\ =\ \partial ^{\,\mu }\ {\omega }^{\,\nu }\
-\ \partial ^{\,\nu }\ {\omega }^{\,\mu }
\end{equation}
\begin{equation}
\ {\vec{R}}_{\rho }^{\ \mu \nu }\ =\ \partial ^{\mu }\ {\vec{\rho}^{\ \nu }}%
\ -\ \partial ^{\nu }\ {\vec{\rho}^{\ \mu }}\ 
\end{equation}
Finally, ${\cal L}_{CC}$ contains the counterterms necessary to eliminate
the divergences from the vacuum polarization. We will discuss this topic
in more detail in the next section .

From Eq. (\ref{L}) one can derive the equations of motion for the nucleons
and mesons. In order to obtain the meson propagators, we have used the
linear response theory around a given ground state of the meson-nucleon
plasma. Our formalism is based on the introduction of Wigner functions. 
This formalism has been described in several papers in connection with 
the nucleon-nucleon interaction in a nuclear medium \cite{GDP94,DP91,DA85}.
We will give here only the main steps relevant to the model adopted above, 
and refer the reader to these papers (and references therein) for a detailed 
description of the method.

The Wigner function for the nucleons is defined by : 
\begin{equation}
F(x,p)=<\hat{F}(x,p)>=tr\left[ {\hat\rho}_{E}\, \hat{F}(x,p)\right]
\label{FW}
\end{equation}
where $\hat{F}(x,p)$ represents the nucleon Wigner operator 
\begin{equation}
\hat{F}(x,p)\ =\ \frac{1}{(2\pi)^{4}} \int d^{4}R \ e^{-ipR} \ \ {\overline{%
\psi}}\,(x+\frac{R}{2}\,)\, \otimes \ {\psi} \,(x-\frac{R}{2}\,)
\end{equation}
and $\,{\hat{\rho}}_{E}$ is the equilibrium density matrix operator. For a
system at a given temperature $T$ allowing for particle number variation,
this is given by the Grand Canonical operator. Given a quantum operator 
$\hat{O}$, the statistical average is defined, as in Eq. (\ref{FW}), by 
\begin{equation}
<\hat{O}>=tr\left[ {\hat{\rho}}_{E}\ \hat{O}\right]
\end{equation}
(the symbol $tr$ means the trace with respect to the quantum states
available to the system).

We assume that equilibrium can be described by the Hartree approximation.
Within this approximation, meson fields are treated as classical fields and
replaced by their statistical averages. These are furthermore restricted by
symmetry properties. For symmetric nuclear matter, only $<\sigma >$ and the
time-like component $<\omega ^{0}>$ survive \cite{[SW86]} (all other meson
mean fields are zero). This means that the nuclear background is described
by the Walecka model. The mean-field values have to be obtained
self-consistently, by solving the implicit equations 
\begin{equation}
\mu _{\sigma }^{2}\ <\sigma >\ =\ \,g_{\sigma }Tr\int d^{4}p\,\ F_{H}(p)
\end{equation}
\begin{equation}
\mu _{\omega }^{2}\ <\ {\omega }^{0}>\ =\,-\,g_{\omega }\,\ Tr\int 
d^{4}p\,\ \gamma ^{0}F_{H}(p)
\end{equation}
Here, $Tr$ stands for the spin-isospin trace. In a spin saturated system, 
parity conservation implies $< \pi >=0$. The value of the $\rho$ 
mean field is related to the difference between neutron and proton 
densities, and since we are working in symmetric nuclear matter, we have
$< \rho >=0$.

The nucleon Wigner function in the Hartree approximation is given by 
\begin{equation}
F_{H}(p)\ =\ (\,\gamma \cdot P\,+\,M\,)\ \left( 
\matrix{
f(p) &  & 0 \cr 
0 &  & f(p) \cr}
\right)  \label{FH}
\end{equation}

The nucleon effective mass is $M=m-g_{\sigma }<\sigma >$, and $P^\mu$ is
defined by $P^{\mu }=p^{\mu
}+g_{\omega }<\omega ^{\mu }>$ . The last matrix in Eq. (\ref{FH})
corresponds to the isospin structure of $F_{H}(p)$ , and \ $f(p)$ is the
relativistic distribution function of nucleons, defined by 
\begin{equation}
f(p)\ =\ \frac{1}{(2\pi )^{3}}\ \delta (P^{2}\,-\,M^{2})\ [\,\Omega
^{+}(p_{0})\,+\,\Omega ^{-}(p_{0})\,-\,H(-P_{0})\,]
\label{reldistrib}
\end{equation}
where $H(x)$ is the Heaviside step function and $\Omega ^{+}(p_{0})$ ($%
\Omega ^{-}(p_{0})$) are the nucleon (anti-nucleon) occupation numbers: 
\begin{equation}
\Omega^{\pm}(p_0)\ =\ H(\pm P_0)\ \ \frac{1}{[\ 1\,+\,
e^{\,\beta\,(\,p_0\,\mp\,\mu\,)\ }]}
\end{equation}
In the latter equation, $\mu $ is the nucleon chemical potential and $\beta
=1/T$ (we take the Boltzmann constant $k_{B}=1$).

The next step in the linear response formalism is made by introducing small
perturbations of the meson fields and the nucleon Wigner function around
their Hartree values. Therefore, one has to replace :

\beq
{\sigma }(x)  & \rightarrow &  <\sigma > + \delta\, {\sigma } (x) 
\nonumber \\
{\omega }^{\mu }(x) & \rightarrow &  <\omega^{\mu }> + 
\delta\, {\omega }^{\mu }(x) 
\nonumber \\
{\vec{\pi}}(x) & \rightarrow &  \delta\, {\vec{\pi}}(x) \nonumber \\
{\vec{\rho}}^{\mu }(x) &\rightarrow & \delta\, {\vec{\rho}}^{\mu }(x) 
\nonumber \\
{F}(x,p) &\rightarrow &  F_{H}(p) + \delta {F}(x,p)
\enq
in the corresponding equations of motion, and consider only terms which are
linear in the perturbations (notice that $<\vec{\pi}>=<\vec{\rho}^{\ \mu
}>=0 $ , as discussed above). One then obtains a system of coupled equations
which can be solved for $\delta F(x,p)$ . When substituted in the remaining
equations, it gives a system of equations for the meson sector. For
symmetric nuclear matter, the resulting system is considerably simplified.
First, the pions and the rho mesons decouple from the rest. Only the $\sigma 
$ and $\omega $ mesons are mixed together. However, the corresponding $%
\sigma +\omega $ equations are the same that appear in the Walecka model 
\cite{GDP94,CHIN77}. Similarly, one reproduces for pions the
well-known results \cite{[SW86],DP91}. For this reason, we will concentrate
here on the rho mesons. In this case, because of isospin symmetry, the
equations of motion are the same for $\rho ^{+}$, $\rho ^{-}$ and $\rho ^{0}$%
, and can be written as : 
\begin{equation}
D^{\mu \nu }(k)\ \delta \rho _{\nu }\equiv \ \left[ \,-k^{\mu }k^{\nu }\ +\
(\ k^{2}\ -\ \mu _{\rho }^{2}\ )\ g^{\mu \nu }\ +\ {\Pi }_{\rho }^{\mu \nu
}(k)\ \right] \ \delta \rho _{\nu }\ =\ 0  \label{D(k)}
\end{equation}
Here, $\Pi _{\rho }^{\mu \nu }(k)$ is the $\rho $-meson polarization tensor
due to NN particle-hole loops. Explicitly,
\beq
\Pi^{\mu\nu}_{\rho} &=& 
\int d^4 p\ Tr \left[ (\gamma .(p-{k\over 2}) +M) 
(g_\rho \gamma^\mu +{f_\rho \over 2 m} \sigma^{\mu\alpha} k_\alpha) 
(\gamma .(p+{k \over 2}) +M) (g_\rho \gamma^\nu -{f_\rho \over 2 m} 
\sigma^{\nu\beta} k_\beta) \right]\ \times \nonumber \\
& & \qquad \qquad  \times \  
{f(p-k/2) - f(p+k/2) \over p.k} 
\enq
(Analytical formulae are given in Appendices A and B). This tensor can be 
decomposed into two terms (we omit Lorentz indices): 
\begin{equation}
\Pi_{\rho }(k) \ = \ \Pi_{\rho }^{mat}(k)\,+\,\Pi_{\rho }^{vac}(k)
\end{equation}

The {\it matter} polarization, $\Pi _{\rho }^{mat}(k)$ vanishes in free
space, i.e. at zero density and temperature. The second, {\it vacuum }term $%
\Pi _{\rho }^{vac}(k)$ , gives a non-zero contribution even in free space.
It contains divergent integrals from which one has to extract a finite
contribution. This will be discussed in the next section.

The $\rho $-meson propagator in the medium $G(k)$ can be obtained by
analyzing the response of the meson field against external nuclear sources.
This procedure has been described in detail in \cite{GDP94}. As showed in
this reference, the resulting propagator is given, in a matrix form, by 
\begin{equation}
G(k)=-\left[ D(k)\right] ^{-1}  \label{G(k)}
\end{equation}
and obeys the Dyson equation (also written matricially). This equation,
graphically represented in Fig. 1, reads 
\begin{equation}
G(k)=G^{0}(k)+G^{0}(k)\Pi_{\rho }(k)G(k)
\end{equation}
where $G^{0}(k)$ is the non-interacting $\rho $-meson propagator in vacuum.
Eq. (\ref{D(k)}) can be analyzed with more detail by choosing a particular
frame with the $z$-axis along the direction of $\vec{k}$ , in such a way
that $k^{\mu }=(w,0,0,q)$. Then, the only non-vanishing components of the
polarization are $\Pi _{\rho }^{00}(k)$, $\Pi _{\rho }^{11}(k)=\Pi _{\rho
}^{22}(k)$,$\Pi _{\rho }^{03}(k)=\Pi _{\rho }^{30}(k)$ and $\Pi _{\rho
}^{33}(k)$. Eq. (\ref{D(k)}) then becomes 
\begin{equation}
\left[ 
\matrix{
D^{00}(k) &  & 0 &  & 0 &  & D^{03}(k) \cr 
0 &  & D^{11}(k) &  & 0 &  & 0 \cr
0 &  & 0 &  & D^{22}(k) &  & 0 \cr
D^{30}(k) &  & 0 &  & 0 &  & D^{33}(k) \cr}
\right] \ \times \left[ 
\matrix{
& \delta \rho_{0}(k) &  \cr 
& \delta \rho_{1}(k) &  \cr
& \delta \rho_{2}(k) &  \cr
& \delta \rho_{3}(k) &  \cr}
\right] \ =\ 0  \label{D(k)frame}
\end{equation}

where 
\begin{eqnarray}
D^{00} (k)\  & = & \ - \ ( \ q^2 \ + \ \mu_{\rho}^2 \ - \ {\Pi}%
^{00}_{\rho}(k) \ ) \\ 
D^{11} (k) \ = D^{22} (k) \  & = & \ - \ ( \ k^2 \ - \ \mu_{\rho}^2 \ - \ {%
\Pi}^{11}_{\rho}(k) \ ) \\ 
D^{33} (k) \  & = & \ - w^2 \ + \ \mu_{\rho}^2 \ + \ {\Pi}^{33}_{\rho}(k) \\ 
D^{03} (k) \ = D^{30} (k) \  & = & \ - w q \ + \ {\Pi}^{03}_{\rho}(k)
\label{D(k)explicit}
\end{eqnarray}

The dispersion relations can now be obtained by equating to zero the
determinant of Eq. (\ref{D(k)frame}), which can be factored out into two
terms : 
\begin{equation}
|D(k)|\ =\ (D^{11} (k))^2 .\ \left| 
\matrix{
D^{00} (k) & D^{03} (k) \cr 
D^{30} (k) & D^{33} (k) \cr}
\right| \ =\ 0
\end{equation}

We obtain two equations, corresponding to {\it transverse} and {\it %
time-longi\-tu\-di\-nal} modes. These are given, respectively, by : 
\begin{equation}
k^{2}\ -\ \mu _{\rho }^{2}\ -\ {\Pi }_{\rho }^{11}(k)\ =\ 0  \label{reldispt}
\end{equation}
and 
\begin{equation}
k^{2}\ -\ \mu _{\rho }^{2}\ -\ k^{2}/q^{2}\ {\Pi }_{\rho }^{00}(k)\ =\ 0
\label{reldispl}
\end{equation}

It is convenient to give these results in a fully covariant form.
The $\rho$ meson polarization can be decomposed as
\beq
\Pi_\rho^{\mu\nu} &=&- \Pi_{\rho T} T^{\mu\nu} -\Pi_{\rho L} L^{\mu\nu} \qquad
\mbox{\rm with} \nonumber \\
T^{\mu\nu} &=& g^{\mu\nu} -{k^\mu k^\nu \over k^2} - {\eta^\mu \eta^\nu
\over \eta^2} \qquad ; \qquad L^{\mu\nu} =  {\eta^\mu \eta^\nu \over \eta^2}
\qquad ; \qquad \eta^\mu = u^\mu - {k.u \over k^2} k^\mu 
\enq
where $u^\mu$ is the 4-velocity of the medium. $\Pi_{\rho T}$ and 
$\Pi_{\rho L}$ are Lorentz scalars representing the transverse and longitudinal
components of the polarization. The dispersion relations read
\beq
k^2 - \mu_\rho^2 + \Pi_{\rho T}=0 \qquad ; \qquad k^2 - \mu_\rho^2 + \Pi_{\rho L}=0 
\enq
The propagator is obtained by inversion of the dispersion relation, and is
given in the general case by
\beq
G_\rho^{\mu\nu} = -G_{\rho L} L^{\mu\nu} -G_{\rho T} T^{\mu\nu} 
+{1 \over \mu_\rho^2} {k^\mu k^\nu \over k^2} \quad ; \qquad
G_{\rho L}= {1 \over k^2 -\mu_\rho^2 + \Pi_{\rho L} }\quad ; \qquad
G_{\rho T}= {1 \over k^2 -\mu_\rho^2 + \Pi_{\rho T}}
\label{covarprop}
\enq
In the referential where the fluid is at rest $u^\mu=(1, \vec 0)$, we have 
the following relations:
\beq
\Pi_{\rho L}=  -\ k^{2}/q^{2}\ {\Pi }_{\rho }^{00}(k)  \qquad ; \qquad
\Pi_{\rho T}=   -\ {\Pi }_{\rho }^{11}(k)
\enq

\section{Vacuum polarization}

In this section we discuss the procedure to extract a finite contribution from
the vacuum polarization tensor. In the case of the $\sigma $ and $\omega $, or 
$\pi$ mesons with a pseudoscalar coupling, the Lagrangian is renormalizable. 
This means that divergences can be eliminated consistently, at all orders, by 
introducing appropriate counterterms in the Lagrangian and imposing some 
physical conditions. This has been discussed in several papers; for a 
discussion within the formalism of Wigner functions see 
\cite{DP91,DA85,[H78]}. As pointed
out in these references, the vacuum polarization gives an important 
contribution and eliminates some of the pathologies that appear in the 
{\it semi-classical approximation} (when all vacuum terms are simply discarded 
\footnote{Moreover, the vacuum contribution depends on the plasma 
thermodynamical state, so that it can not be simply subtracted in a fully 
consistent way.}). Therefore, one would like to conserve vacuum effects. Here, 
however, there is an important difference when considering the rho mesons. 
As a matter of fact, due to the derivative coupling appearing in Eq.
(\ref{LI}), the Lagrangian becomes non-renormalizable, which implies that 
the counterterms needed to compensate the infinities at a given order of 
approximation will not be valid at higher orders. Nevertheless, it is still 
possible to eliminate the divergences {\it at a given order} by the procedure 
described below. 

The method proceeds by dimensional regularization and the introduction of
a counterterm Lagrangian.  

The vacuum contribution arises from the last term $-H(-P_0)$ in Eq. 
(\ref{reldistrib}). 
Since the only tensors at our disposal in vacuum are $g^{\mu\nu}$
and $k^\mu k^\nu$, and $\Pi^{\mu\nu}_{\rho\ vac}$ should be orthogonal
to $k^\mu k^\nu$, it must be of the form:
\begin{equation}
\Pi _{\rho ren}^{\mu \nu }(k)\ =\ -\,Q^{\mu \nu }\ \frac{\displaystyle I(k)}{%
\displaystyle 4\pi }  \label{polvac}
\end{equation}
where 
\begin{equation}
Q^{\mu \nu }(k)\ =\ {\displaystyle\frac{k^{\mu }k^{\nu }}{k^{2}}\ -\ g^{\mu
\nu }\ }  \label{tensorQ}
\end{equation}
After performing the trace, a straightforward calculation leads to
\beq
\Pi_{\rho\ vac}(k) &=& -{16 \over 3} {g_\rho^2\over (2 \pi)^3} {\cal I}_1
-{4 \over 3} {g_\rho^2 \over (2 \pi)^3 } (2 M^2 k^2 + k^4) {\cal I}_2 -
{8 \over 3} \left( {f_\rho \over 2 m} \right)^2 {k^2 \over (2 \pi)^3}
{\cal I}_1 -{2 \over 3} \left( {f_\rho \over 2 m} \right)^2 
{k^4 \over (2 \pi)^3} (8 M^2 +k^2) {\cal I}_2 \nonumber \\
& & -{8 M k^4 \over (2 \pi)^3} \left( {f_\rho \over 2 m} \right) 
g_\rho {\cal I}_2 
\enq
The integrals ${\cal I}_1$ and ${\cal I}_2$ diverge. These divergences can
be extracted by the procedure of dimensional regularization
\beq
{\cal I}_1 & = & \int d^4 p\ \delta(p^2-M^2)\, H(-p_0) \quad; \qquad 
{\cal I}_1^{reg}=\pi M^2 \left[ -{1 \over \epsilon} + \ln \left( {M\over m}
\Lambda_1 \right) \right] \\ 
{\cal I}_2 & = &  \int d^4 p\ {\delta(p^2 -M^2)\, H(-p_0) \over (p.k)^2 -k^4/4}
\quad ; \qquad
{\cal I}_2^{reg} = -{2 \pi \over k^2} \left[ -{1 \over \epsilon}
+ \ln \left( {M \over m} \Lambda_2 \right) +\theta(k^2,M^2) \right]
\label{divint}
\enq
$\theta(k^2,M^2)$ is a known finite function given in Appendix B.
Here, $\epsilon$ is an infinitesimal quantity  and $\Lambda_1$, $\Lambda_2$ 
are arbitrary finite constants. Therefore, there appear the following 
divergences in $\Pi_{\rho\ vac}$ when $\epsilon \rightarrow \infty$:
\beq
{1 \over \epsilon}\quad {\rm in\ the\ } g_\rho^2 {\rm \ term}\, ,\qquad
{M \over \epsilon}\quad {\rm in\ the\ } {f_\rho \over 2 m} g_\rho{\rm \ term}
\, ,
\qquad {M^2 \over \epsilon}\quad {\rm and}\quad {k^2 \over \epsilon}\quad 
{\rm in\ the\ } \left( {f_\rho \over 2 m} \right)^2 {\rm \ term}
\label{divergences}
\enq
At the {\sl one-loop level}, it is possible to compensate the divergences by 
adding the following counterterms to the Lagrangian:
\begin{equation}
{\cal L}_{CC}^{\rho }=\left\{ (A\ +\ B\ {\sigma }\ +\ C\ {\sigma 
}^{2}\ )\ \ {\vec{R}}^{\mu \nu }\,\cdot \,{\vec{R}}_{\mu \nu }\ +\ D\
(\partial _{\alpha }\ {\vec{R}}_{\mu \nu })\cdot (\partial ^{\alpha }\ {\vec{%
R}}^{\mu \nu })\right\}
\end{equation}
in which $A$, $B$, $C$, $D$ are constants to be determined by the
renormalization procedure. In order to obtain the finite (renormalized)
vacuum polarization, we have followed the same subtraction scheme as we used
for the other mesons \cite{DP91}. Namely, we first obtain the rho field 
equation from the generalized Euler equation\footnote{The $D$ counterterm 
makes this generalization necessary}
\beq
{\partial {\cal L} \over \partial \rho_\nu} - \partial_\mu \left( 
{\partial {\cal L} \over \partial(\partial_\mu \rho_\nu )} \right) +
\partial_\alpha \partial_\mu \left( {\partial {\cal L} \over 
\partial(\partial_\alpha \partial_\mu \rho_\nu )}\right)  =0
\enq
with the full Lagrangian including the counterterms. After linearizing 
and Fourier transforming, the dispersion equation (\ref{D(k)}) reads
\beq
\left[  -k^\mu k^\nu + (k^2 -\mu_\rho^2) g^{\mu\nu} + \Pi_\rho^{\mu\nu}
+(A +B \sigma +C \sigma^2 +D k^2)(k^\mu k^\nu -k^2 g^{\mu\nu})
\right] \delta \rho_\nu = 0
\enq
Now the vacuum rho polarization can be made finite if $A,B,C,D$
consist of an infinite part cancelling the divergences we have pointed 
out in Eq. (\ref{divergences}), and a finite part which we now proceed to
determine on the basis of physical arguments. We will have
\beq
\Pi_{\rho\, ren}^{\mu\nu} &=& \Pi_{\rho\, vac}^{\mu\nu}  
+(A +B \sigma +C \sigma^2 +D k^2) (k^\mu k^\nu -k^2 g^{\mu\nu}) \nonumber \\
&=& \left\{ {g_\rho \over 3 \pi^2} \left[ \ln {M\over m} + \theta 
+ {2 M^2 \over k^2} (\theta -1) \right] + \left( {f_\rho \over 2 m} \right)^2 
{1 \over 6 \pi^2}  \left[ 6 M^2  \ln {M\over m} + 8 M^2 \theta 
+ k^2 (\ln {M\over m} + \theta) \right] \right. \nonumber \\
& & \left. + {2 M \over \pi^2} \left({f_\rho \over 2 m}\right) g_\rho 
\left[ \ln {M\over m} + \theta \right]  - \left( \alpha + \beta\, \sigma 
+ \gamma\, \sigma^2 + \delta k^2 \right) \right\}
\left\{ k^2 g^{\mu\nu} -k^\mu k^\nu \right\}
\enq
with the decomposition
\beq
A &=& \alpha + {g_\rho^2 \over 3 \pi^2} \biggl[ -{1 \over \epsilon} 
+ \ln \Lambda_2 -{1 \over 2} \biggr]\! +\! {2 m \over \pi^2} 
\left({g_\rho\, f_\rho \over 2 m}\right)\! \biggl[-{1 \over \epsilon} 
+ \ln \Lambda_2 -{1 \over 2} \biggr]\! +\! \left( {f_\rho \over 2 m} \right)^2\! 
{m^2 \over 3 \pi^2}  \biggl[ -{3 \over \epsilon} + 4\, \ln \Lambda_2 
-\ln \Lambda_1 - 2 \biggr] \nonumber \\
B &=& \beta - {2 g_\sigma \over \pi^2} \left({g_\rho\, f_\rho \over 2 m}\right) 
\left[-{1 \over \epsilon} + \ln \Lambda_2 -{1 \over 2} \right] 
- \left( {f_\rho \over 2 m} \right)^2 {2 m g_\sigma\over 3 \pi^2}  
\left[ -{3 \over \epsilon} + 4\, \ln \Lambda_2 -\ln \Lambda_1 - 2 \right] 
\nonumber \\
C &=& \gamma + \left( {f_\rho \over 2 m} \right)^2 {g_\sigma^2\over 3 \pi^2}  
\left[ -{3 \over \epsilon} + 4\, \ln \Lambda_2 -\ln \Lambda_1 - 2 \right] 
\nonumber \\
D &=& \delta + \left( {f_\rho \over 2 m} \right)^2 {1 \over 6 \pi^2}  
\left[ -{1 \over \epsilon} + \ln \Lambda_2 -{1 \over 2} \right] 
\enq 
In order to determine the finite constants $\alpha,\beta,\gamma,\delta$, 
some physical constraints have to be imposed. The standard \cite{CHIN77} 
requirements are that:
\vskip 0.2cm \noindent
\newline\hskip 0.5cm  * (1) 
The vacuum polarization has to vanish in free space, when the rho meson is on
its mass shell (here, ``shell''=$\{ k^2=\mu_\rho^2, M=m, \sigma=0 \}$): 
\beq \displaystyle{{\Pi_{\rho\ vac}}_{| {\rm shell}}}=0, 
\label{ren1}
\enq
\newline\hskip 0.5cm  * (2) 
The field equation can be written in the standard form 
by defining effective coupling constants $g_{\rm eff}$ and $f_{\rm eff}$. 
Requiring that the 
effective coupling be equal to the bare one in vacuum leads to the condition
\beq 
\displaystyle{{\partial \Pi_{\rho\ vac} \over \partial k^2}_{| {\rm shell}}}=0
\label{ren2}
\enq
\newline\hskip 0.5cm  * (3) 
The new couplings $\sigma R_{\mu\nu} R^{\mu\nu}$, $\sigma^2 
R_{\mu\nu} R^{\mu\nu}$, that we had to introduce in order to subtract 
the infinities, should not appear in the dispersion relation on the rho 
meson mass shell, leading to the conditions
\beq
\displaystyle{{\partial \Pi_{\rho\ vac} \over \partial \sigma}_{| 
{\rm shell}}} =0 \label{ren3}
\enq 
\newline and 
\newline\hskip 0.5cm  * (4) 
\beq
\displaystyle{{\partial \Pi_{\rho\ vac} 
\over \partial \sigma^2}_{| {\rm shell}}}=0
\label{ren4}
\enq
\newline\noindent This amounts to four relations to determine the four
constants $\alpha, \beta, \gamma, \delta$. 

Alternative sets of conditions can be imposed, leading to different expressions
of the vacuum polarization. In particular, we can recover the renormalization
scheme suggested by Shiomi and Hatsuda \cite{[SH94]} as a special case of
our general formalism. We have also tried these alternative schemes in our 
calculations in order to investigate the influence of the treatment of the 
vacuum. As we will see in section (\ref{effmass}), the various schemes 
predict a completely different behavior of the in-medium rho meson mass 
as density increases. 

The peculiarities of each scheme, outline of derivation and corresponding 
analytical expressions are given in Appendix B. They can be subdivided into 
two classes: The first class of schemes comprises those which are 
designed not to introduce new couplings with the $\sigma$ mesons (conditions 
(3) and (4) of the method outlined above). This is the case of the scheme 
presented here, labelled ``scheme 1'' and of a very similar one labelled 
``scheme 2''. With a different point of renormalization, the scheme presented 
in the case of the $\sigma$ meson in the classical paper of Kurasawa and 
Suzuki \cite{[KS88]}
also belongs to this class. We applied it to the $\rho$ meson and labelled 
the result ``scheme 6''.  The second class of renormalization schemes
preserves the structure of the regularized expression, introducing counterterms
$\alpha + \beta \sigma + \gamma \sigma^2$ so that they can be factorized
into $\alpha M^2/ m^2 = \alpha( m-g _\sigma \sigma)^2/m^2 $. As a result, 
conditions (3) and (4) cannot be fulfilled anymore. We called the scheme 
obtained in this way ``scheme 3''. The method of Shiomi and Hatsuda 
(scheme 5) and a related one used by Sarkar {\it et al.} \cite{Sarkar98} 
(``scheme 4'') belong to this class.

The advantages and caveats of the respective renormalization schemes 
will be further discussed in a forthcoming work\cite{DMP00}.
\vskip 0.5cm

\section{One-boson exchange potential in the medium.}

In this section we calculate the one-boson exchange potential obtained after
RPA summation. As mentioned in the previous section, when the background is
symmetric nuclear matter, rho mesons (and pions as well) decouple from the
other mesons, whereas $\sigma $ and $\omega $ mesons are coupled together.
Therefore, the total potential shows the following structure : 
\begin{equation}
V=V^{\sigma +\omega }+V^{\pi }+V^{\pi }
\end{equation}
where $V^{\sigma +\omega }$,$V^{\pi }$ and $V^{\pi }$ are obtained by $%
\sigma +\omega $, one-pion and one-rho exchange, respectively. Our method
follows a similar procedure to the construction of the one-boson exchange
potential in vacuum \cite{M89}. The essential difference is that free meson
propagators are replaced by in-medium propagators. Also, the external
nucleon lines correspond to in-medium spinors, as they arise by solving the
nucleon Dirac equation in the Hartree approximation. It is important to
mention, however, that one arrives to the above result by computing the
energy associated to a pair of interacting nucleons inside the plasma, a
result which ensures that no double counting has been made.
The method has been given in detail in \cite{GDP94}, and 
applied to the calculation of $V_{\sigma +\omega }$.

\bigskip

As in the free case, we expand the resulting potential in powers of $\frac{p%
}{M}$, where $p$ represents the momenta of the external nucleons, and keep
only terms of the order $\sim \left( \frac{p}{M}\right) ^{2}$. This has the
advantage to simplify the spin structure of the nucleon-nucleon potential.
For degenerate nuclear matter, one has $p\sim p_{F}$, where $p_{F}$ is the
nucleon Fermi momentum. If the density is close to saturation, the next term
in the expansion would be a small correction of the order $\left( \frac{p}{M}%
\right) ^{4}\sim \left( \frac{0.3}{0.7}\right) ^{4}$. Of course, if density
is much larger, higher order terms in the above expansion should be taken
into account.

In constructing the one-rho potential one has to take into account the
diagram shown in Fig. 2, where the double line represents the in-medium
rho-meson propagator, given by Eqs. (\ref{G(k)} -- \ref{D(k)explicit})
and Eq. (\ref{covarprop}).
The amplitude corresponding to this diagram is given by 
\begin{eqnarray}
{\cal M}_{\rho } & = & -  \left\{ \left[ \, \chi _{1^{\prime }}^{\dagger
}\, \overline{u}(\vec{p^{\prime }}_{1},s_{1}^{\prime })\,\right] \ \left(
\,i\,g_{\rho }\,\gamma ^{\mu } - \frac{\displaystyle f_\rho}{\displaystyle 2m}
\, \sigma ^{\mu \beta }(p_1\,^{\prime }-p_1)_{\beta}\right) \ \tau_1^a\ 
\left[ \, u(\vec{p}_{1},s_{1})\, \chi _{1}\ \right] \,\right\} \nonumber \\ 
& &  \delta _{ab}\ G_{\mu \nu }(k)\ \ \left\{ \left[ \, \chi
_{2^{\prime }}^{\dagger }\, \overline{u}(\vec{p^{\prime }}_{2},s_{2}^{\prime
})\,\right] \left( \,i\,g_{\rho }\,\gamma ^{\nu } - \frac{\displaystyle %
f_{\rho }}{\displaystyle2m}\, \sigma ^{\nu \alpha }(p_{2}\,^{\prime
}-p_{2})_{\alpha }\right) \, \tau_2^b \,\left[ \ u(\vec{p}_{2},s_{2})\
\chi _{2}\ \right] \,\right\}
\label{Mrho}
\end{eqnarray}
In this equation, $\chi _{1},\chi _{2}$ stand for the quantum states of the
initial nucleons with four-momenta ${p}_{1}$ and ${p}_{2}$. They have
associated spinors $u(\vec{p}_{1},s_{1})$ and $u(\vec{p}_{2},s_{2})$ .
Similarly, the prime indicates the corresponding magnitudes for final
nucleons. The isospin operators $\tau _{1}^{a},\tau _{1}^{b}$ ($a,b=+,-,0$)
label the associated rho charged fields. From Eq. (\ref{Mrho}) we obtain the
one-rho exchange potential, by using the procedure outlined above. In the 
center-of-mass frame (CM) of the nucleons\footnote{We consider here the case 
where the rest frame of the background fluid coincides with the center of 
mass of the collision. For the effects due to a relative velocity between 
both frames, the reader is referred to \cite{JDA-LM-PLB}}, the potential 
is given by: 
\begin{eqnarray}
V_\rho (\vec q,\vec Q) &=& \Bigg\{ g_\rho^2 \  
\left[\ -G_{\rho L} \left( 1+{\vec Q^2 \over 2 M^2} -{\vec q{}^2 \over 8 M^2} 
+ {i \over 2 M^2} (\vec q \wedge \vec Q) . \vec S  \right)
 \right. \nonumber \\
&&\left. \qquad \ \ \,
+\, G_{\rho T} \left ( -{\vec Q^2 \over M^2} - {i \over M^2} 
( \vec q \wedge \vec Q). \vec S + (\vec \sigma_1 . \vec \sigma_2) 
{\vec q{}^2  \over 4 M^2} - {(\vec \sigma_1 . \vec q) (\vec \sigma_2 . 
\vec q) \over 4 M^2} \right) \right] \nonumber \\
&& +\, 4 M\, g_\rho \left({f_\rho \over 2 m}\right) \
\left[\ -G_{\rho L} \left( -{\vec q{}^2 \over 4 M^2} + {i \over 2 M^2} 
( \vec q \wedge \vec Q ) . \vec S \right) \right. \nonumber \\
& & \left. \qquad \qquad \qquad \quad \ \ 
+\, G_{\rho T} \left( -{i \over 2 M^2} ( \vec q \wedge \vec Q) . \vec S +
 (\vec \sigma_1 . \vec \sigma_2) {\vec q{}^2  \over 4 M^2} 
- {(\vec \sigma_1 . \vec q) (\vec \sigma_2 . \vec q) \over 4 M^2} 
\right) \right] \nonumber \\
&& +\, 4 M^2 \left({f_\rho \over 2 m}\right)^2 \left[\ G_{\rho T} 
\left( (\vec \sigma_1 . \vec \sigma_2) {\vec q{}^2  \over 4 M^2} 
- {(\vec \sigma_1 . \vec q) (\vec \sigma_2 . \vec q) \over 4 M^2} 
\right) \right]\ \Bigg\}
\quad \vec{\tau}_{1}\cdot \vec{\tau}_{2}  \label{Vrho}
\end{eqnarray}
In the latter equation, $\vec{\sigma}_{i}$ ($\vec{\tau}_{i}$) represent the
spin (isospin) Pauli matrices for the two ($i=1,2$) interacting nucleons.
In the chosen frame, initial nucleons have momenta $\vec{p}$ and $-\vec{p}$ ,
whereas final nucleons are assumed to have momenta $\vec{p}^{\ \prime }$ and -$%
\vec{p}^{\ \prime }$ . We have introduced the notations : 
\beq
&& \vec{q} =\vec{p}-\vec{p}^{\;\prime }  \qquad ; \qquad 
\vec{Q} =\left( \vec{p}+\vec{p}^{\;\prime }\right) /2 \qquad ; \nonumber \\
&& G_{\rho L} (0,\vec q) = {-1 \over q^2 +\mu_\rho^2 - \Pi_\rho^{00}(0,\vec q)} 
\qquad ; \qquad
G_{\rho T} (0,\vec q) = {-1 \over q^2+\mu_\rho^2+\Pi_\rho^{11}(0,\vec q)} 
\enq
Next we will construct the in-medium (RPA summed) nuclear potential in
configuration space. This is done by Fourier transformation of the
corresponding momentum-space magnitudes. Before proceeding, we include a
phenomenological form factor in the nucleon-meson vertices. This amounts
to making the replacements : 
\begin{eqnarray}
g_{\alpha } &\rightarrow &g_{\alpha }\cdot {\cal F}_{\alpha }(q)  \nonumber
\\
f_{\rho } &\rightarrow &f_{\rho }\cdot {\cal F}_{\rho }(q)  \nonumber \\
{\cal F}_{\alpha }(q) &=&\frac{\Lambda _{\alpha }^{2}-\mu _{\alpha }^{2}}{%
\Lambda _{\alpha }^{2}+q^{2}}
\end{eqnarray}
($\alpha =\sigma ,\omega ,\pi ,\rho $) in all previous equations in momentum
space. After some algebra, the configuration space potential can be
decomposed in the following way \cite{M89} : 
\begin{eqnarray}
V(\vec{r}\,)\  & =\  & \ V_{c}(r)\,-\,\ {\displaystyle\frac{1}{2}}\ \left( \
\nabla ^{2}\,V_{NL}(\vec{r}\,)\ +\ V_{NL}(\vec{r}\,)\nabla ^{2}\ \right)
\nonumber \\ 
&  & \,+\,V_{LS}(r)\ \vec{L}\cdot \vec{S}\,+\,V_{SS}(r)\ \vec{\sigma}%
_{1}\cdot \vec{\sigma}_{2}\ +\ V_{T}(r)\ S_{12}
\label{V(r)}
\end{eqnarray}
with the following notations : $\vec{L}$ is the angular momentum operator, $%
\vec{S}=\frac{1}{2}\left( \vec{\sigma}_{1}+\vec{\sigma}_{2}\right) $ is the
total spin of the nucleons and $S_{12}$ the tensor operator. In this way, $%
V_{c}(r)$ is the central part of the potential, $V_{LS}(r)$ is the
spin-orbit, $V_{SS}(r)$ is the spin-spin, and $V_{T}(r)$ is the tensor
potential. In addition to the spin structure, one has to include a factor $%
\vec{\tau}_{1}\cdot \vec{\tau}_{2}$ in the pion and rho contributions
(omitted here). The second  term in Eq. (\ref{V(r)}) is non-local.
Its contribution will depend on the explicit form of the
two-nucleon wave function and, therefore, can not be directly plotted. For
our analysis in the next section we will not consider this term.

Each one of the pieces in Eq. (\ref{V(r)})  has contributions from
different meson exchanges ($\sigma +\omega $ sector, $\pi $ or $\rho $).
Their expressions can be found in Appendix C.

\section{Results for the rho-meson propagation}

\subsection{Choice of parameters}

We will analyze the behavior of the in-medium potential as density changes,
at zero temperature. Our calculations include the vacuum and matter 
polarization in the meson propagators. Since the vacuum contribution does not 
vanish at finite $\vec q$ in general, we cannot take directly the parameters 
available in the litterature. Instead, we must perform a fit anew for
each one of renormalization schemes described in Appendix B. We have fitted
the value of the coupling constants $g_\sigma$, $g_\omega$, $g_\rho$, 
$f_\rho$ and cutoff parameters $\Lambda_\sigma$, $\Lambda_\omega$, 
$\Lambda_\pi$, $\Lambda_\rho$ in such a way that, at zero density, the model 
reproduces as close as possible the Bonn 'Potential B' \cite{M89} (table A.3 
in this reference). The value of the pion coupling $g_\pi$ was kept constant 
since it is well known from experimental data.
The values of meson masses are taken directly from this reference. However,
we do not include the $\delta $ and $\eta $ mesons in the fitting procedure.
Since the Bonn potential has been adjusted to reproduce low-energy nucleon
scattering data, we expect that our model can give a reasonable description
to these experimental data. We have chosen to make this fit in the region
from $0.5$ to $2.5$ {\it fm} for all mesons. The resulting parameters are 
given in Table 1, and the values of the obtained chi-squared for the 
different pieces of the potential appear in Table 2.

\begin{table}[htbp]
\begin{center}
\begin{tabular}{|c|c|c|c|c|c|c|c|c|}
\hline
 & $g_\sigma$ & $g_\omega$  & $g_{\rho}$ & $f_\rho/g_\rho$ & 
$\Lambda_{\sigma}$ & $\Lambda_\omega$ & $\Lambda_\pi$ & $\Lambda_\rho$ 
 \\
\hline
\hline
Bonn B & 10.6 & 17.55 & 3.36 & 6.1 & 1900 & 1850 & 1700 & 1850 \\
\hline
\hline
set 1A & 8.10 & 16.267 & 4.216 & 4.974 & 1630.55 & 1562.19 & 1179.48 
& 1881.93 \\
set 1B & 7.55 & 15.817 & 3.390 & 5.115 & 1558.22 & 1550.12 & 729.40 
& 1365.54 \\
\hline
set 2A & 7.83 & 15.783  & 3.718 & 4.996 & 1390.92 & 1493.97 & 976.30 
& 1797.30 \\
set 2B & 7.59 & 16.091 & 3.882 & 4.319 & 1532.25 & 1521.51 & 690.48 
& 1112.22  \\
\hline
set 3A & 8.54 & 15.910 & 3.862 & 5.811 & 1080.65 & 1441.00 & 1212.30 
& 1638.73  \\
set 3B & 8.53 & 16.377 & 5.710 & 4.547 & 1329.54 & 1500.82 & 1150.89
& 1260.05 \\
\hline
set 4A & 7.22 & 14.624 & 3.303 & 7.033 & 1540.38 & 1629.70 & 1144.72 
& 1375.11 \\
\hline
set 6A & 7.57 & 14.809 & 3.636 & 5.731 & 1626.54 & 1590.76 & 1125.17
& 1711.54 \\
\hline
\end{tabular} \\
\caption{Values the coupling constants and cutoffs which adjust
the Bonn potential B \cite{M89} for each renormalization scheme}
\end{center}
\end{table}

\begin{table}[htbp]
\begin{center}
\begin{tabular}{|c|c|c|c|c|}
\hline
 & $\chi^2(V_C)$ & $\chi^2(V_{SS})$  & $\chi^2(V_T)$ & $\chi^2(V_{LS})$  
 \\
\hline
\hline
set 1A & 0.096 & 0.026 & 0.014 & 0.078 \\
set 1B & 0.136 & 0.232 & 0.019 & 0.179 \\
\hline
set 2A & 0.078 & 0.049 & 0.056 & 0.172 \\
set 2B & 0.123 & 0.182 & 0.061 & 0.203 \\
\hline
set 3A & 0.026 & 0.039 & 0.035 & 0.142 \\
set 3B & 0.038 & 0.065 & 0.031 & 0.082 \\
\hline
set 4A & 0.079 & 0.079 & 0.054 & 0.210 \\
\hline
set 6A & 0.087 & 0.034 & 0.069 & 0.191 \\
\hline
\end{tabular} \\
\caption{$\chi^2$ values corresponding to the parameter sets given in Table 1,
for each component of the $NN$ potential: central ($C$), spin-spin ($SS$), 
tensor ($T$) and spin-orbit ($LS$)}
\end{center}
\end{table}

In these tables, the number labelling each set is determined by the 
corresponding renormalization scheme (see Appendix B). The letter A means 
that the best fit for all components of the potential was looked for. 
As will be seen in the next section, the dispersion relation of the $\rho$ 
meson may display heavy meson modes and, more annoyingly, zero-sound modes 
for too high values of the $\rho$ meson cutoff. Therefore we performed
the fit once again, but now with the constraint that a lower cutoff is used
for the $\rho$ meson in order to avoid the appearance of these 
spurious branches. The letter B corresponds such fits, and due to 
this restriction they come with slightly higher $\chi^2$ values.

We do not give parameter sets for renormalization scheme 5, because
this scheme was constructed by Shiomi and Hatsuda in such a way 
that the vacuum polarization vanishes identically (for all $q$)
in the vacuum. In this case, the expression of Machleidt \cite{M89}
is recovered exactly, therefore there is no need of readjusting the 
parameters and the original values of Machleidt (Bonn potential B) 
may be used.

\subsection{Rho meson dispersion relations}

With the above values of the meson parameters, we have first
performed a numerical study of the rho-meson\footnote{The analysis of 
the $\sigma $ , $\omega $ and $\pi $ dispersion relations has been 
performed in \cite{DP91}.} dispersion relations defined by Eqs. 
(\ref{reldispt}), (\ref{reldispl}). Vacuum appears through the 
renormalization scheme discussed in Section 3. We use the 
renormalization scheme described in that section (scheme 1 of 
Appendix B), as well as alternative schemes described in Appendix B.
A sample of the resulting branches are plotted in Figures 3 to 5 for 
longitudinal and transverse modes, for various values of the nucleon 
Fermi momentum  $p_{F} =$ 0.3, 0.4, 0.5 (all magnitudes are given 
in units of the nucleon bare mass $m$). In our model, this correspond 
to densities $\rho_0$, 2.37 $\rho_0$ and 4.63 $\rho_0$ with $\rho_0$ 
being the nuclear saturation density. The variables $\omega$ and $q$ are 
defined by $k=(\omega,0,0,q)$, as in Section 2. 

We at once observe that the dispersion relation strongly depends on the 
chosen renormalization scheme, both directly, and indirectly since it
conditions the choice of parameters at the time of fitting the potential 
(see Table 1). 
Let us first discuss the general features of the transverse mode
for parameter sets 1A to 6A, which were obtained by fitting the
$NN$ potential without imposing restrictions on the allowed range for
each parameter. 

In the timelike region, besides the normal 
branch, we also have two heavy meson branches in the general case. 
This kind of meson branches have been obtained in other models of QHD, 
and originate from vacuum effects \cite{DP91,DA85}. Such meson modes 
are present at densities corresponding to $p_{F}\geq 0.3$ for all 
renormalization schemes, with the exception of parameter set 4A,
corresponding to the renormalization scheme of Sarkar \cite{Sarkar98}. 
It must be noted, however, that this parameter set has a value of the 
$\rho$ meson cutoff appreciably lower than other A-sets. As will be 
discussed later, a smaller value of the cutoff has the effect of reducing 
or eliminating the heavy meson branches.

At high density, there appears moreover a zero sound branch in the spacelike 
region. This latter branch could have especially strong (and unpleasant)
effects on our results for the potential, since it appears as a pole in 
the propagator. We found such a branch for renormalization schemes belonging
to both classes considered in this work: in the ``increasing rho mass'' 
class, with parameter set 2A at $p_F > 0.4$ and also for the ``decreasing
rho mass'' renormalization scheme of Shiomi and Hatsuda at $p_F > 0.6$.
Nevertheless, we will see in the sequel that this pole can be eliminated 
by reducing the value of the $\rho$ meson cutoff.  

The longitudinal modes have a similar structure in the timelike 
region, with a normal branch and two heavy meson branches. In the spacelike 
region no zero sound branches were found.

In order to eliminate the spurious branches, it was possible to find 
a second series of parameter sets with stronger cutoffs for the $\rho$ meson,  
at the cost of a slightly degraded quality of the fit. We called the
resulting sets 1B, 2B and 3B. This permits to eliminate the zero sound
branch at all densities investigated (up to $p_F$=2) and to reduce or 
eliminate the heavy meson branches. It was moreover checked that the
choice of a lower cutoff parameter does not affect appreciably 
the position of the normal branch.

The dispersion relations for parameter set 1B are shown in Figure 3.
A moderate cutoff $\Lambda_\rho$=1365 MeV was applied. The zero sound branch 
is now removed, but we still have heavy meson branches. The left panel
displays the transverse modes. The right panel compares the transverse
(thin line) to the longitudinal (thick line) modes for the normal branch.
At $q=0$ the transverse and longitudinal modes coincide; for finite
$q$ they differ only slightly. Finally, we note that the intercept of 
the normal branch with the $q=0$ axis goes to higher frequencies with 
increasing $p_F$ (or densities). This is related to the fact that the 
effective mass of the $\rho$ meson increases with this renormalization 
schemes, as discussed in next section.

The dispersion relations for parameter set 2B are shown in Figure 4, now
with a strong cutoff $\Lambda_\rho$=1112 MeV. Only the normal branch 
remains for the transverse mode. The longitudinal modes are also 
appreciably cleaned, with only a persistent heavy meson mode at high 
momentum transfer. A still lower cutoff would eliminate it, but
would spoil the accuracy of the fit of the $NN$ potential.

In Figure 5 we show the transverse dispersion relation obtained with 
a renormalization scheme of the second (``decreasing rho mass'') class. 
The parameter set used is 3B, with a reasonable $\rho$ cutoff 
$\Lambda_\rho$=1260 MeV, resulting in a clean dispersion relation:
there only remains the normal branch. Moreover, the quality of the fit 
remains very good. We do not show the longitudinal modes since we found 
that longitudinal and transverse modes are indistinguishable.
We note that the intercept of the normal branch with the $q=0$ axis is lower 
at finite density than in vacuum, and slightly increases from $p_F$=0.3 
to $p_F$=0.5, corresponding to the fact that the rho meson effective mass 
with this renormalization scheme first decreases with density and reaches 
a minimum around $p_F=0.35$ (see next section).

We also programmed and plotted the dispersion relations obtained with
schemes 4 to 6, but chose not to display them here . Scheme 6 
(``Kurasawa-Suzuki'') was found to be essentially similar to other schemes 
of the same class (1 and 2) but more problematic when we tried to remove 
the sound mode and heavy meson branches.
For the dispersion relation obtained with the scheme 5 of Shiomi and 
Hatsuda \cite{[SH94]} with the parameters of Machleidt Bonn potential B, 
we have again a normal branch  and two heavy meson branches 
in the longitudinal as well as in the transverse modes. Moreover, a tiny 
zero sound branch also appears in the transverse mode  at high enough 
densities ($\rho > 7\ \rho_0,\ p_F < \sim 0.6$). We performed a fit
with a lower value of $\Lambda_\rho$ and found a parameter set 
which behaves essentially like case 3B. For the scheme of Sarkar {\it et al.} 
(scheme 4) we obtain only the normal branch for both transverse and 
longitudinal modes. This renormalization scheme is free from heavy meson 
branches and zero sound modes. This is due in part to the fact that the fit 
to the potential in vacuum requires in this case typically lower rho meson 
cutoffs of the order of 1300 MeV. Still we found this scheme more robust 
than the previous ones, since no such branches were found even for extreme 
values of parameters adjusting the potential.

\subsection{Rho meson effective mass}
\label{effmass}

From the dispersion relations one can obtain the in-medium mass $\mu_{eff}$,
defined as the solution $\mu _{eff}=\omega$ of the dispersion relations
Eq. (\ref{reldispt}), (\ref{reldispl}) at $\vec q=0$. The transverse and 
longitudinal modes yield the same effective mass, since the polarizations 
are equal in this limit. In Figs. 6-8 we show how $\mu_{eff}$ evolves, 
in units of the vacuum mass $\mu _{\rho }$, as $p_{F}$ grows (remember 
that $p_{F}/m=0.3$ corresponds to saturation density). In particular, we 
wish to investigate the role of the vacuum on its behavior, in connection
with the claim of Shiomi and Hatsuda \cite{[SH94]} that the vacuum 
contribution solves this issue.

We first show in Figure 6 the effective mass obtained by keeping only 
the matter contribution in the rho meson polarization (thus neglecting 
the vacuum term). This figure was obtained with the parameter set
of Machleidt potential B (see table 1). If we would limit ourselves
to a moderate density and frequency, it would appear that the rho meson mass
is slightly increasing. A closer inspection reveals, however, that the
structure of the dispersion relation is more complex and has 3 solutions
for $p_F < 0.246$ (that is, a density $\rho < 0.5 \rho_0$), corresponding
to anomalous branches intermingled with the normal one.
At higher density, only the higher branch survives. This type of behavior
is in fact well documented. It was first observed in the case of the
$\omega$ meson by Lim and Horowitz \cite{LH89} and also appears
in the case of the pion \cite{JDAprivate}. Contrarily to a rather widespread 
belief, it is not cured by the introduction of a stronger form factor.
As a matter of fact, we redraw on the same figure the rho mass for a low
cutoff (1 GeV) and see that the problem, although less acute, is still
present. Moreover, we cannot choose the cutoff as we please, since it was
obtained from a fit of the nucleon-nucleon potential in the vacuum.

In the case of the $\sigma$, $\omega$ and $\pi$ mesons, the situation is
largely improved by introducing the vacuum term. There, a lower normal
branch cleanly separates from higher heavy meson branches. Without
any cutoff, two heavy meson branches appear at zero density and merge 
and disappear at high density. The heavy meson branches can thereafter 
be removed by the application of a reasonable cutoff. As we will see, 
the same occurs in the case of the rho meson. 

In figure 7, the on-shell renormalization schemes 1 and 2 of Appendix B 
were used. The parameters were chosen so as to obtain the best fit of 
the nucleon-nucleon potential in each case (parameter sets 1A and 2A 
of table 1). We also plotted on the same figure the resulted obtained 
by adapting the scheme of Kurasawa and Suzuki for the $\sigma$, $\omega$ 
mesons to this case (parameter set 6A). In this figure and the next one, 
only the mass corresponding to the lower normal branch is represented. 
There also exist two branches at higher masses (typically 3.5 $\mu_\rho$ 
and 7 $\mu_\rho$) which merge and disappear at very high density, which we 
do not show here. It is a common feature of all these renormalization 
schemes that the lower branch gives an in-medium mass which is larger 
than in vacuum. In all cases, the increase is quantitatively similar for 
all the three schemes, and somewhat stronger for scheme 6 of Kurasawa
and Suzuki. On the other hand, the heavy meson branches largely depend 
on the value of the cutoff, so that this cutoff can be used to remove 
these branches altogether. As an example, we give in Table 1 a
parameter set which makes it possible both to obtain a satisfactory
fit of the potential and to remove the heavy meson branches (set 2B).

On the other hand, if one makes use of the renormalization scheme 
proposed by Shiomi and Hatsuda in \cite{[SH94]}, one then obtains 
a rho mass which decreases with density. We have re-derived the formulae 
given in this reference using our own methods (scheme 5 of Appendix B) 
and checked that our result is in agreement with theirs. The resulting 
effective mass is plotted in Figure 8. As advertised, the rho meson mass 
decreases until $p_F=0.35$ (or $\rho=1.6 \rho_0$), and then slowly increases. 
We note that there also appear heavy meson branches. We plotted on 
the same figure the result obtained from a similar scheme used by 
Sarkar {\it et al.} \cite{Sarkar98} (scheme 4 of Appendix B). While 
Shiomi and Hatsuda subtract the vacuum at all momenta $k$, Sarkar 
{\it et al.} set the vacuum to zero at $k^2=\mu_\rho^2$ in a way similar 
to our condition Eq. (\ref{ren1}). The normal branch obtained from
scheme 4 is quantitatively very similar to the result of Shiomi 
and Hatsuda.

These results are tantalizing since they are in agreement with the
currently accepted concept of a decreasing rho meson mass based
on other models and experimental data. 
Now, in order to interpret the contradiction which appears between 
schemes 1,2,6 on one hand, and 4,5 on the other hand, we have developped 
a further scheme (scheme 3). As further explained in Appendix B, the
introduction of counterterms $(A + B\, \sigma +C\, \sigma^2)$ does not
preserve in general the original structure of the diverging term
proportional to $M^2= (m-g_\sigma \sigma)^2$. It can be
required that this $M^2$ structure be preserved by dropping conditions
(\ref{ren3},\ref{ren4}). The resulting effective mass is also plotted
on Figure 8, and we see that it actually decreases as well. 

The idea of a decreasing rho mass in the medium was popularized by the 
Brown and Rho scaling conjecture \cite{BR91} as a result of chiral symmetry 
restoration. It allows to parametrize the in-medium mass of vector mesons as
\begin{equation}
\frac{\mu _{eff}}{\mu _\rho}\approx 1-(0.18\pm 0.05)\frac \rho {\rho _0}
\end{equation}
as a function of the density $\rho $ (written in units of the nuclear
saturation density $\rho _{0}$). 
For example, as a consequence of QCD sum rules, Hatsuda and Lee 
\cite{[HL92]} or Leinweber and Jin \cite{Leinweber} obtain a decreasing 
mass. The use of QCD sum rules, however, was criticized for containing 
some uncertainties and inconsistencies with chiral perturbation theory 
\cite{[TC94]}. Another caveat is the parametrization of the spectral
function used in these calculations \cite{LPM98,KKW97}, which further
reduces the predictive power of QCD sum rules arguments. Finally, 
recent QCD sum rules calculations with an improved vacuum subtraction 
\cite{Nyff00} further contributed mitigating the simple Brown-Rho 
decreasing mass picture.

Other calculations, based on chiral approaches (see {\it e.g.} \cite{Oset00},
and \cite{RW99} for a review) or quark-meson coupling models \cite{[ST97]} 
also predict a slight lowering of the rho meson mass. In chiral models including 
a $\rho\pi\pi$ interaction piece (see also the brief discussion in this work, 
section \ref{mesonloops}), the rho mass is only slightly modified, the important 
effect there is the broadening of its spectral function. It is interesting to note 
that, depending on the renormalization scheme, the position of the maximum of 
the spectral function is shifted up- or downwards depending on the choice of 
the renormalization procedure (\cite{[HFN93]} {\it vs.} \cite{klingl}).
The theoretical situation is far to be settled, and there is a possibility 
for the rho mass to increase, rather to decrease, due to medium effects 
(see also {\it e.g.} \cite{[Pi95],[Song]}). 

There are a number of experiments in which it has been claimed that one can 
extract information about the vector mesons in a nuclear medium. In heavy-ion
collisions experiments as HELIOS-3 \cite{HELIOS} and CERES \cite{CERES}, 
the excess of dilepton production at invariant masses lower than the bare 
rho meson mass might be explained in the framework of the vector dominance 
model by assuming a dropping of the in medium rho-meson mass 
\cite{ChR96,LK94,KB96}. This interpretation is sometimes called, for short, the
B/R (Brown-Rho) scenario. However, it has been pointed out that keeping a 
constant rho mass and introducing a medium modification of its width 
\cite{[CEK95],RCW97} could also explain the dilepton excess. 
Friman and Pirner \cite{[Fri97]} have also argued that the contribution
of higher resonances ($^*$N(1720)) to the rho-meson self energy in matter 
plays an important role in shifting the strength to lower invariant masses.
The `broadening scenario' sometimes receives the name of R/W (Rapp-Wambach
\cite{RCW97}) scenario. 

For some time, measurements of polarization-transfer experiments with
polarized protons seemed to favor the dropping-mass hypothesis as well 
\cite{McC92,BW94,UG94}. New analyses of the results with a better treatment
of relativistic effects, however, concluded that the data could be explained
without this assumption \cite{Toki}.
 
In summary, the problem of how medium effects will modify the rho-meson
properties is still not well settled, both from the theoretical and
experimental points of view, although the possibility that its mass will be
smaller at higher densities seems to be more favored. Future experiments,
like HADES at GSI, will hopefully help clarifying this issue.

In our hadronic model, we have seen that the inclusion of the vacuum 
contribution largely affects the meson propagation in matter. Vacuum 
renormalization still allows for a increasing, as well as for a 
decreasing meson mass, and does not by itself settle the issue. 
If we believe that the rho meson mass should be decreasing, then a 
renormalization scheme which preserves the original dependence in the 
nucleon effective mass should be chosen. But then, in order to be 
coherent, the same procedure has to be used for the $\sigma$, $\pi$ 
and $\omega$ mesons as well \cite{DMP00} \footnote{Preliminary studies 
indicate however that this would spoil the behavior of the $\sigma$ and $\pi$ 
effective masses \cite{DMP00}.}.

In view of this situation, we have made some comparisons using the two 
classes of renormalization schemes described above, each one of them 
giving a different behavior of the rho meson inside the nuclear medium.

\section{Results for the RPA effective potential in the medium.}

We shall now analyze the different components of the in-medium potential in
position space, obtained in section 4, as density changes at $T=0$ . As
discussed in that section, the mixing terms of the $\rho $-meson
polarization with other mesons vanish if one considers symmetric nuclear
matter, a circumstance which will allow us to consider first the
contribution to the potential of this meson alone. Later, we shall add the
contributions of the $\sigma $, $\omega $ and $\pi $ mesons.

In performing the present analysis, it is useful to make the following
division in the distance variable $r$, as was made in \cite{GDP94}:

a) In the {\it short-range} region $0\leq r\leq 1$ fm, relativistic effects
(beyond the quadratic approximation considered above) become essential.
Also, in this distance range the extended structure of the nucleon must be
properly taken into account. The introduction of form factors allows to
simulate this effect for not-too-short distances, such as $0.5\leq $ $r\leq
0.8$ fm. For shorter distances, calculations from our model will not be
reliable.

b) The {\it intermediate region} will be defined as the distance range $%
1\leq r\leq 2$ fm. Within this range, the vacuum potential becomes
quantitatively modified by medium and vacuum effects, as compared to the
free-space potential.

c) In the {\it long-range} region : $r\ge 2$ fm, matter polarization
dominates, and one can observe qualitatively new features.

\subsection{Rho-meson exchange}

On the left upper panel of Fig. 9, we show the contribution of one-rho meson 
exchange to the central component of the potential. We compare the 
results obtained for the in-medium potential at $p_{F}/m=0.4$ 
(corresponding to a density $\rho= 2.37 \rho_0$) to the free-space 
potential (dotted curve), which was calculated setting $p_{F}=0$. 
Vacuum polarization effects are kept in all cases. We made the 
calculation for three types of renormalization schemes, one of the first 
``increasing rho mass'' class (scheme 1 with parameter set 1B), 
and two of the ``decreasing rho mass'' class (scheme 5=Shiomi-Hatsuda 
with Machleidt's Bonn B parameters and scheme 3=ours with parameter set 3A).

From this comparison, we observe that the use of a screened interaction 
in matter can bring appreciable modifications to the potential 
described by the exchange of free mesons in the vacuum. In particular,
in the long-range region, the interaction is qualitatively different 
from vacuum. Instead of the usual exponential damping, at non-zero 
density the potential becomes oscillatory  due to the presence of 
{\it Friedel and Yukawa oscillations}. This behavior is obtained
whatever renormalization procedure is chosen.

Friedel oscillations arise because the analytically-continued 
matter polarization shows branch cuts starting at $q=\pm 2p_{F}\,$ in the 
complex $q$-plane. This phenomenon was first discovered for a QED plasma 
\cite{FR52} and was also evidenced in the case of a QCD plasma \cite{KA88}. 
The corresponding analysis for a nucleon plasma in different models was 
performed in \cite {DPS89,GDP94L,GDP94}.
On the other hand, Yukawa oscillations appear when the
analyti\-cally-con\-ti\-nued boson propagator has a pole (in the complex 
$q$-plane) away from the real and imaginary axis. Such a phenomenon has been
found so far for a nucleon plasma in the one-pion exchange approximation,
and for a quark-gluon plasma when one-gluon exchange is considered
\cite{DPS98}. 
In order to separate Friedel from Yukawa effects for rho mesons, 
one needs to perform the analytical continuation of the meson propagator 
and to study the evolution of the Yukawa pole as density and temperature 
evolves. This will be the subject of a future work.

These oscillating phenomena can have consequences if one
goes beyond the Hartree approximation, by including the polarization
contribution into the ground-state energy. In the case of Friedel
oscillations, it has been found that a periodic-density configuration, with
period equal to the characteristic period of Friedel oscillations, has a
lower energy than a constant-density configuration \cite{Pri90}.
This can be interpreted as a transition to a spatially-structured
configuration.

For a fixed renormalization scheme, we obtained that the repulsive 
core extends farther when density increases. On the other hand, for
a given value of the density, the first minimum occurs earlier for
renormalization schemes of the first ``increasing rho mass'' class
than for schemes of the second class. In the long-range zone, the
period of oscillation is seen to be the same for all renormalization
schemes, and depends only on the density, as expected from the
physical origin of Friedel oscillations.

Matter polarization dominates the potential qualitative
features at large distances. The choice of a renormalization scheme can,
however, introduce some quantitative changes on these features. The amplitude 
of the oscillations depends indirectly on the renormalization scheme.
For example, to each scheme, there corresponds a parameter set with
different values of the couplings and cutoffs. The amplitude can be enhanced 
by the presence of a dip in the dispersion relation of the transverse 
modes, indicating the proximity of a zero sound mode, even when no pole 
actually appears. On the other hand we found that the appearance of such 
a branch depends on the renormalization procedure, the second class of 
renormalization schemes being less liable to the appearance of zero sound 
modes. Nevertheless, we would like to stress that the oscillations themselves 
are not a consequence of possible zero sound modes, since we found them also 
in the renormalization scheme of Sarkar, where no zero sound was obtained 
even for extreme choices of the parameters and density.

Similar features can be observed in the remaining components (spin-spin,
tensor and spin-orbit) of the potential. They are plotted in the three 
remaining panels of Fig. 9). They all show an oscillatory behavior.
In all cases, the amplitude of oscillations increases with density. 

\subsection{Combined-meson potential}

We will now present some results which are obtained by adding the
contribution of the $\sigma +\omega $ sector and the $\pi $ exchange to the
potential given above\footnote{The contribution from $\delta $ and $\eta $ 
mesons is not included here since they introduce only a small modification 
to the vacuum potential of \cite{M89}}. For the central component there 
is no pion contribution, and one has : 

\begin{equation}
V_{c}(r)=V_{c}^{\sigma +\omega }(r)+V_{c}^{\rho }(r)
\end{equation}

The result is plotted in Fig. 10 in vacuum (solid curve) and at finite 
densities $p_F=$ 0.3, 0.4, 0.5 (dashed, dot-dashed and dotted lines 
respectively). In this figure the renormalization scheme 1 was chosen 
with parameter set B. The free-space potential (which includes 
vacuum polarization with on-shell renormalization, scheme 1) has a potential 
well with a minimum at $r\sim 1.5$ fm. At finite density, the position of the
first minimum is displaced towards shorter distances: at saturation density 
($p_F=0.3$) it corresponds to $r\sim 1.25$ fm, whereas for $p_F=0.5$ it
is located at $r\sim 0.95$ fm. There also appear 
secondary minima which depth can be significant. In the example chosen, at 
$p_F=0.5$ ($\rho=\ 4.63\, \rho_0$), the second minimum is situated at 
$r \sim 2.1 fm$ with a depth of -8 MeV. The right panel of Fig. 10 
focusses on the long-range oscillatory behavior. Here it can be seen
that the period of the oscillation decreases and its amplitude increases 
with increasing density.

We also studied the spin-spin, tensor and spin-orbit components of the 
potential. The results are not shown here; basically the same features
as already mentioned in the case of the rho component were observed.

\section{Mechanisms reducing the amplitude of oscillations}

In this section we discuss the mechanisms which could suppress
the oscillations found in the in-medium NN potential.

We will first investigate the contribution of meson loops to the 
$\rho$-meson self energy. As is well known, the spectral function of 
the rho meson acquires an important contribution of the $\rho\pi\pi$ 
loop (see {\it eg.} \cite{[HFN93]}). A meson with a mass distribution,
such as the $\sigma$ or the $\rho$, could smear away some features in the 
potential by merging contributions with different ranges. As we will
see, the effect on the potential is negligible. For consequences
on the dilepton production in the vector dominance model, we refer 
the reader to the vast litterature devoted to this subject (see {\it
eg.} \cite{[HFN93],klingl,urban,cassingbratk} and references therein).

A second and more effective mechanism is the rounding off of the edge
of the Fermi momentum distribution function introduced by short-range 
correlations or by a finite temperature. This will be investigated 
in the second part of this section.

\subsection{Meson loops}
\label{mesonloops}

We have up to now studied the contribution of nucleon-hole loops to the 
$\rho$ meson polarization. On the other hand, the $\rho$ meson also 
couples to the pion, a fact which is at the origin of the decay width 
of the $\rho$ meson.
Let us add to the Lagrangian a piece
\beq
{\cal L}_{\rho\pi\pi}=  g_{\rho\pi\pi} (\partial^\mu \vec \pi \times \vec \pi) 
. \vec \rho_\mu +{1 \over 2} g_{\rho\pi\pi}^2 (\vec \rho^\mu \times \vec \pi).
(\vec \rho_\mu \times \vec \pi)
\enq
At the first order of the linear expansion method of \cite{DA85},
meson loops do not appear. It would be necessary to go to higher
orders in the cluster expansion, so that meson-meson correlators appear. 
We would like here to estimate the correction brought by 
the $\rho\pi\pi$ polarization loop. From Green function techniques,
one obtains \cite{Sarkar98,[HFN93],[Asakawa]}
\beq
\Pi_{\rho\pi\pi}= i g_{\rho\pi\pi}^2 \int {d^4 p \over (2 \pi)^4} 
(2 p-k)^\mu (2 p -k)^\nu D_\pi (p) D_\pi (p-k) -2 i g_{\rho\pi\pi}^2
\int {d^4 p \over (2 \pi)^4}  g^{\mu\nu} D_\pi(p)
\enq
In general, the pion propagator also includes medium effects. In order to keep 
things simple however, we will keep here the free pion propagator
$D_\pi^0(k)= 1/(k^2-\mu_\pi^2+i \epsilon)$.

The coupling constant $g_{\rho\pi\pi}$ is determined by fixing the
imaginary part of the polarization on the $\rho$ meson mass shell
and in the rest frame to its experimental value
\beq
\Gamma_{\rho\pi\pi} &=& - {1 \over \mu_\rho} {\cal I}m \Pi_{\rho\pi\pi} 
                       (\omega=\mu_\rho,\vec q=0) \
                    = \ {g_{\rho\pi\pi}^2 \over 48 \pi \mu_\rho^2} 
                        (\mu_\rho^2 -4 \mu_\pi^2)^{3/2}
\enq
Putting $\Gamma_{\rho\pi\pi} \simeq 151.5$ MeV, one finds 
$g_{\rho\pi\pi}^2/(4 \pi) \simeq $ 2.9
The real part of the polarization can be expressed in terms of the
divergent integrals (\ref{divint}) 
\beq
{\cal R}e\, \Pi_{\rho\pi\pi}^{\mu\nu}= 
- {2 \over 3} {g_{\rho\pi\pi}^2 \over (2 \pi)^2}
\left[ 2 {\cal I}_1 + k^2 (\mu_\pi^2 - {k^2 \over 4}) {\cal I}_2 \right]
\left( g^{\mu\nu} - {k^\mu k^\nu \over k^2} \right)
\enq
After performing dimensional regularization, 
\beq
{\cal R}e\, \Pi_{\rho\pi\pi}^{\mu\nu}= 
- {2 \over 3} {g_{\rho\pi\pi}^2 \over (2 \pi)^2} 
\left( g^{\mu\nu} - {k^\mu k^\nu \over k^2} \right)
\left[ ({k^2 \over 4} - \mu_\pi^2)
\theta(k^2,\mu_\pi^2) -{k^2 \over \epsilon} \right] 
\enq
The infinity is cancelled by a counterterm of the form $A_{\rho\pi\pi}
R^{\mu\nu} R_{\mu\nu}$, that is to say, the renormalization proceeds
exactly as in the unambiguous case of the $NN$ loop contribution to
the $\omega$ meson. The finite result is:
\beq
{\cal R}e\, \Pi_{\rho\pi\pi}^{\mu\nu}= 
- {2 \over 3} {g_{\rho\pi\pi}^2 \over (2 \pi)^2} 
\left( g^{\mu\nu} - {k^\mu k^\nu \over k^2} \right)
\left[ {\mu_\pi^2 \over \mu_\rho^2} (\mu_\rho^2 -k^2) 
\left( 1 -\theta(\mu_\rho^2,\mu_\pi^2) \right) + 
(\mu_\pi^2-{k^2 \over 4}) \left( \theta(\mu_\rho^2,\mu_\pi^2) 
- \theta(k^2,\mu_\pi^2)\right) \right]
\enq

The imaginary part is given by
\beq
{\cal I}m\, \Pi_{\rho\pi\pi}^{\mu\nu}=  
              {g_{\rho\pi\pi}^2 \over 48 \pi } k^2 
              \left(1 -4 {\mu_\pi^2\over \mu_\rho^2} \right)^{3/2} 
              \theta(k^2-4 \mu_\pi^2)
\enq

In the calculation of the potential, the imaginary part will not contribute
since the polarizations entering the formulae are to be taken at $\omega=0$
while the imaginary part should be an odd function of $\omega$ so that the
Onsager relations be fulfilled. 
The real part of the polarization yields a small shift to the dispersion 
relation. We have checked that its effect on the potential is negligible. 
At the level of approximation considered here, the shift is $q$-dependent 
but does not depend on thermodynamical conditions. It would if we introduced 
the modification of the pion propagator by $N \Delta$ loops or a bath of 
thermal pions (see {\it eg.} \cite{urban}).

\subsection{Effect of short range correlations}

A second mechanism which can attenuate the long range oscillatory behavior
of the in-medium potential is the rounding-off of the momentum distribution
by short-range correlations of the Brueckner type. As commented in the 
introduction, a fully consistent calculation at the level of parquet
approximation would be required. Here, in order to estimate the
order of magnitude of this effect, we will simply introduce such a momentum 
distribution instead of the Fermi-Dirac one in the calculation of the 
polarizations. 

Among the results available in the literature we chose two: a non relativistic 
Brueckner-Hartree-Fock calculation of Baldo {\it et al.} starting from a
separable Paris interaction \cite{Baldo90}, and a Dirac-Brueckner-Hartree-Fock 
calculation by de Jong and Malfliet \cite{dJM91}. It seems more appropriate 
to use the calculation of de Jong and Malfliet since it is fully relativistic 
and was obtained starting from the Bonn potential. Moreover, these authors 
checked explicitly that the thermodynamical consistency be optimally fulfilled.
On the other hand, comparison with experimental data seems to favor the
larger depletion and smaller discontinuity found in nonrelativistic
calculations such as that of Baldo {\it et al.}, so that we 
also present results with this distribution.

We found that the data published by de Jong and Malfliet was well reproduced
by the following fit at saturation density:
\beq
F_{DBHF}(k)= \left\{ \matrix{ 0.91-0.055\bigg( \displaystyle{k\over p_F} \bigg)
-0.07 \bigg( 1-\displaystyle{k\over p_F} \bigg)
\log\bigg(1-\displaystyle{k \over p_F}\bigg) & {\rm if}\ k<p_F \nonumber \\
 \displaystyle{0.035\over k/p_F-0.94}\ \exp \bigg(-1.2\, 
\displaystyle{k\over p_F} \bigg) & {\rm if}\ p_F < k < 1.5 p_F \nonumber \\
0.3693\ \exp\bigg( -2.4\, \displaystyle{k \over p_F}\bigg) & 
{\rm if}\ k > 1.5 p_F \nonumber \\  }\right.
\enq
For the data of Baldo {\it et al.}, we used
\beq
F_{NRBHF}(k)= \left\{ \matrix{0.79-0.13 \bigg( \displaystyle{k \over p_F} 
\bigg) -0.19 \bigg(1.-\displaystyle{k\over p_F} \bigg) \log\bigg( 
1-\displaystyle{k \over p_F} \bigg)   & {\rm if}\ k<p_F \nonumber \\
16.\ \exp \bigg(-4.5\, \displaystyle{k \over p_F} \bigg)  & 
{\rm if}\ p_F < k < 1.5 p_F \nonumber \\
0.2127\ \exp \bigg( -1.6\, \displaystyle{k \over p_F} \bigg)   
& {\rm if}\ k > 1.5 p_F \nonumber\\  }\right.   
\enq

We now calculate the polarizations at vanishing temperature as
\beq
\Pi_{BHF}(k)= \int_0^\infty d^3 p\ {\partial \Pi^{(T=0)}_{FD} (k,p) \over \partial p} \
F_{BHF}(p) 
\enq
where $\Pi^{(T=0)}_{FD}$ is the polarization which would have been obtained with a 
Fermi Dirac step function $\theta(p_F-p)$ and is given in Appendix A. 
Next we introduce these
polarizations in the expression of the potentials given in Appendix C.
Figure 11 compares the central potential as obtained with parameter set 1B
at vanishing temperature and saturation density for the Fermi-Dirac
momentum distribution (dashed line), the relativistic DBHF distribution of 
de Jong and Malfliet (full line) and the nonrelativistic BHF distribution of
Baldo {\it et al.} (dotted line). It is seen that the oscillations are damped
by Brueckner correlations as was to be expected. The effect is however
not very severe and the oscillations persist. 

\subsection{Effect of finite temperature}

A similar damping of the oscillations is produced by a non-vanishing
temperature. We found that the Brueckner modification of the momentum
distribution had the same effect as a temperature of $\sim$ 10 MeV 
with a Fermi-Dirac distribution. It can be seen on Figure 12 that
Friedel oscillations are rather robust against temperature and are
not washed out for temperatures as high as 30 MeV.

As a by-product, we obtained the behavior of the $\rho$ effective mass as
a function of temperature for all renormalization schemes. At saturation
density, it was found that $\mu_{eff}$ (calculated as in section \ref{effmass})
slightly decreases with temperature for renormalization schemes of the 
first class 1, 2 and 6, whereas it slightly increases for renormalization
schemes of the second class 3, 4 and 5. At four times the saturation density,
 $\mu_{eff}$ was found to decrease for all renormalization schemes.
The results are well described by a linear law for temperatures
$T \in [0-100]$ MeV: 
\beq
{\mu_{eff} \over \mu_\rho} (\rho,T) = {\mu_{eff} \over \mu_\rho} (\rho,0)
 \left( 1-a {T \over m} \right)
\enq
with 
\begin{center}
\begin{tabular}{lcll}
$a=0.4$   & for $\rho=\rho_0$     & renorm. scheme= & 1 or 2 \\
$a=0.35$  &   $\quad$   "   "     &  $\quad$ " $\qquad$  "  & 6  \\
$a=-0.12$ &   $\quad$   "   "     &  $\quad$ " $\qquad$  "  & 3, 4 or 5 \\
          &                       &                 &       \\
$a=1.5$   & for $\rho=4\, \rho_0$ & renorm. scheme= & 1 or 2 \\
$a=0.8$   &   $\quad$   "   "     &  $\quad$ " $\qquad$  "  & 6  \\
$a=0.55$  &   $\quad$   "   "     &  $\quad$ " $\qquad$  "  & 3, 4 or 5 \\
\end{tabular}
\end{center}

\section{Discussion and Conclusions}

In this paper we have discussed RPA effects on the nucleon-nucleon potential
obtained within the one-boson exchange approximation in symmetric nuclear
matter at zero temperature. The model we used for our calculations includes 
$\sigma $ , $\omega $ , $\pi $ and $\rho $ mesons interacting with nucleons
via Yukawa couplings. The Wigner function technique and linear response 
analysis were used to obtain the meson propagation in medium. The lowest 
order of this scheme is the mean-field approximation. When vacuum effects are
properly renormalized, it becomes the Hartree approximation. The next order
of this approximation has been shown to be equivalent to Green's function 
calculations of the meson propagators at one-loop order\cite{DP91}.

In our calculations, medium effects appear in two ways. First, in the
nucleon legs of the one-boson exchange diagram the nucleon effective mass
appears. Secondly, the meson propagators are calculated including 
nucleon-hole loops and vacuum polarization effects.

It is legitimate to ask whether one should consider other loops involving 
mesons or resonances \cite{[Fri97]}. Only a brief discussion 
concerning pion-pion loops was given, since several thorough studies 
exist in the litterature \cite{[HFN93],[Asakawa],urban}. Loops involving 
resonances can be expected to have a smaller but sizeable contribution 
of similar characteristics to the nucleon loop taken into account in this 
work. A serious study with renormalization would, however, go beyond
the scope of this paper. 

In symmetric nuclear matter, rho mesons decouple from other mesons.
Therefore, it is possible to study one-$\rho $ exchange alone and simply add
the contribution of $\sigma $ , $\omega $ and $\pi $ to the nucleon-nucleon
potential at the end. We have first studied the {\it dispersion relations}
which describe the propagation of rho-mesons in matter as density changes.
One observes the presence of two types of branches : a {\it normal branch},
which is the analogous of the free-space mass-shell condition, and several 
{\it heavy-meson branches}. The latter appear as a consequence of vacuum
effects, in all the QHD meson models that have been investigated
\cite{DP91,DA85,[DH84]}. Here, due to the derivative coupling of the rho-mesons
to the nucleons, the Lagrangian is non-renormalizable. Yet, it is possible
to extract a finite contribution from the vacuum at each order in the
cluster expansion\footnote{We use the word {\it renormalization} in
a wide sense to describe this procedure, in spite of the non-renormalizability 
of the model.}. This procedure, however, contains some arbitrariness in 
the case of non-renormalizable Lagrangians.

In view of this, we have compared the results arising from two different
classes of renormalization schemes. The first one is analogous to the 
one usually employed in previous works \cite{CHIN77,[KS88],DPS89} and is
designed to remove, as far as possible, the spurious new couplings to the
sigma field which have to be introduced in the counterterm Lagrangian. 
The renormalization procedure suggested in \cite{[SH94]} has also been 
used for comparison. We found that it belongs to a second class of 
procedures, where the new couplings to the sigma field are not minimized, 
but instead the original structure of the expression of the vacuum
contribution is preserved. 

By using the first scheme, we find that the effective mass of rho-mesons
grows with density, a feature which is disfavored by most of present
theoretical approaches and by the interpretation of dilepton production
data in heavy-ion collisions. The second method
gives the opposite behavior : the in-medium rho-meson mass drops as density
grows, more in agreement with present ideas. However, it is recognized 
that large uncertainties are contained both in the theoretical calculations 
\cite{[Pi95],[TC94]} and the analysis of experiments mentioned above 
\cite{[CEK95],RW99,cassingbratk}. We have kept for these reasons these two 
schemes in our calculations.

Next, we investigated the RPA nucleon-nucleon potential obtained by the
exchange of one rho meson. At short and intermediate distances, it shows
a repulsive behavior. In the long-range it contains new qualitative features, 
as compared to the free-space potential. It becomes oscillatory, with an 
amplitude which decreases with distance, due to the combination of 
Yukawa-like and Friedel oscillations. The appearance of this oscillatory 
behavior might give rise to a new phase of dense matter, characterized 
by a spatially structured density distribution. We have verified that 
the qualitative features within this distance range are not sensitive 
to the renormalization scheme used. One then expects that they are to 
be found in other models describing rho-meson propagation in dense 
nuclear matter.

When other mesons are added into the potential, we find a similar behavior.
The potential is repulsive at short distances and attractive at intermediate
distances. In the long-range, $r\geq 2$ fm, we find again an oscillatory
behavior in all components of the potential, with the same origin as above.
The amplitude of these oscillations increase with density. 

The robustness of these oscillations against several possible damping 
mechanisms has been tested. We first discussed the influence of pion-pion 
loops. In the {\it spacelike} zone explored by the potential, we found that 
it should not have a decisive impact on the results presented. This does 
not enter in contradiction with the large body of publications for which 
the spectral function is needed in the {\it timelike} region.

We also investigated the amount of smoothing of the Friedel oscillations 
to be expected from modifications to the Fermi surface. In particular, 
we discussed the effect of the temperature and short-range correlations. 
We performed a calculation with a momentum distribution rounded off by 
short range correlations of the Brueckner type, and found they have
the same consequences as a temperature of about 10 MeV, namely, that
the oscillations experiment a moderate damping, and are seen discernible. 
They only disappear for temperatures higher than 30 MeV.

One has to exert some care when considering the applicability to the 
above results to actual nuclear-matter calculations. The RPA summation 
concerns only a class of many-body diagrams, so that the addition of 
different diagrams, such as ladder diagrams, might modify our results
(one would expect, however, that in the long-distance range the RPA results
would hold). The obtained potential can not be used naively to perform 
more elaborated calculations (such as ladder resummation) since it would 
result in double counting of diagrams. Our results rather have to be 
interpreted as an indication of a new qualitative effect that one might find. 

To summarize, screening effects can appreciably modify the nature of
nuclear interactions in a medium. Vacuum effects also compete to modify the
interaction, so that it is important to design a procedure which incorporates
them in a consistent way. However, the appearance of singular behaviors, as
Friedel and Yukawa-like oscillations seems to be a rather general property
of the nuclear interaction.

\section*{Acknowledgments}

This work has been partially supported by the Spanish DGES Grant PB97-1432 and
CICYT AEN96-1718.

\section*{Appendix A: Polarization, matter part}

\setcounter{equation}{0}

We give here the complete formulae for the matter part of the rho-meson 
polarization tensor in symmetric nuclear matter. The vacuum contribution 
is given in Appendix B. This tensor can be written as : 
\begin{equation}
\Pi _{\rho }^{\mu \nu }(k)\ =\ -\,\int d^{4}p \ I^{\mu \nu }(p,k)\ J(p,k)
\label{polmatconint}
\end{equation}
where 
\begin{eqnarray}
I^{\mu\nu}(p,k) &=& \qquad g_\rho^2\ \bigg[ \left( 4\, M^2 +k^2 -4\, 
     p^2 \right) g^{\mu\nu}  -2\, k^\mu k^\nu +8\, p^\mu p^\nu \bigg] 
     \nonumber \\
     & & \!\!\!\!\!\!\!\!
     + \left( { f_\rho \over 2 m} \right)^2 \bigg[ \left( 4\, M^2 k^2 +
     k^4 +4\, p^2 k^2 -8\, (p.k)^2 \right) g^{\mu\nu} 
     -\left( k^2 +4\, M^2 -4\, p^2 \right) k^\mu k^\nu  \nonumber \\
     & &  \qquad \qquad
     +8\, (p.k) \left( k^\mu p^\nu + p^\mu k^\nu \right) 
     -8\, k^2 p^\mu p^\nu \bigg] \nonumber \\
     & & \!\!\!\!\!\!\!\!
     +  \left( { f_\rho \over 2 m} \right) g_\rho
     \bigg[  8\, M  \left( k^2 g^{\mu\nu} -k^\mu k^\nu \right) \bigg] 
\end{eqnarray}
In this equation, $M$ is the nucleon effective mass, and $m$ its free mass.
We have introduced : 
\begin{equation}
J(p,k)=\frac{f(p+k/2)-f(p-k/2)}{k\cdot p}
\end{equation}
The nucleon distribution functions, as they arise from the relativistic
Hartree approximation, are the same for protons and neutrons (in symmetric
nuclear matter). They are given by 
\begin{equation}
{\ \displaystyle f(p) = \frac{\displaystyle \delta (\, p^2-M^2 \,)}{%
\displaystyle (2 \pi)^3} \, \,\left[ \,\frac{H ( p^0\, ) }{e^{\,\beta \,
(p^0-\mu_{eff})} + 1} \,+ \, \frac{H ( p^0\, ) }{e^{\,\beta \,
(p^0+\mu_{eff})} + 1} \,- \,H (-p^0 \, ) \,\right] }  \label{f(p)}
\end{equation}
Here, $\mu_{eff}=\mu +g_{\omega }<\omega^{0}>$ is the nucleon effective
chemical potential, $\mu $ is the true chemical potential and $\beta $ is
the inverse temperature. The time-like mean field value $<\omega^{0}>$ of
the $\omega $ meson is also calculated within the Hartree approximation.
Finally, $H(x)$ is the Heaviside step function.

At zero temperature, it is possible to perform the integrals in Eq. (\ref
{polmatconint}) analytically. Moreover, in order to obtain the
nucleon-nucleon potential, within the approximations discussed in this work,
we only need to calculate the matter polarization on the $k^{0}=0$ axis. In
this case, one gets more compact expressions. We give here these formulae.
By defining $q=|\vec{k}|$, $\varepsilon =\sqrt{M^{2}+(q/2)^{2}}$ and the
nucleon Fermi energy $E_{F}=\sqrt{M^{2}+p_{F}^{2}}$, we can introduce the
notations 
\begin{equation}
A(q)\, =\, {\displaystyle \frac{q-2 p_F}{q +2 p_F}} \ ,\quad  
B(q)\, =\, {\displaystyle \frac{q\,E_F\,+\,2\,p_F \ \varepsilon} 
{q\,E_F\,-\,2\,p_F \ \varepsilon}}
\end{equation}
One finds, after some algebra:  

\begin{eqnarray}
\Pi_{\rho L}(0,q) &=&\Pi_\rho^{00}(0,q) \nonumber \\
   &=& {g_\rho^2 \over  \pi^2} \left[ 
   - {4 \over 3} E_F p_F +{q^2 \over 3} \ln\left( {E_F +p_F \over M} \right)
   -{E_F \over 2} \left( q-{4 \over 3} {E_F^2 \over q} \right) \ln A(q)
   -{\varepsilon \over 2} \left( q-{4 \over 3} {\varepsilon^2 \over q} \right) 
   \ln B(q) \right] \nonumber \\
   & & +\left( {f_\rho \over 2 m} \right)^2 {q^2 \over \pi^2} \left[ 
   -{1 \over 3} E_F p_F + \left( M^2 -{q^2 \over 6} \right)  \ln\left( 
   {E_F +p_F \over M} \right) +2 {E_F \over q} \left( {E_F^2 \over 3} -M^2
   \right) \ln A(q) \right.  \nonumber \\
   & & \left. \qquad \qquad \qquad
     +2 {\varepsilon \over q} \left( {\varepsilon ^2 \over 3}
    -M^2 \right) \ln B(q) \right] \nonumber \\
   & & +\left( {f_\rho \over 2 m} \right) g_\rho {2 M q^2 \over \pi^2} \left[
    \ln\left( {E_F +p_F \over M} \right) -{E_F \over q} \ln A(q) -
   {\varepsilon \over q} \ln B(q) \right] \\
\end{eqnarray}
for the longitudinal part, and 
\begin{eqnarray}
\Pi_{\rho T}(0,q) &=& - \Pi_\rho^{11}(0,q) \nonumber \\
   &=& -{g_\rho^2 \over 2 \pi^2} \left[ 
    {2 \over 3} E_F p_F -{2 \over 3} q^2 \ln\left( {E_F +p_F \over M} \right)
   -E_F \left( 2 {\varepsilon^2 \over q} -{2 \over 3} {E_F^2 \over q} -q 
   \right) \ln A(q) -\varepsilon \left( {4 \over 3} {\varepsilon^2 \over q} 
   -q \right) \ln B(q) \right] \nonumber \\
   & & -\left( {f_\rho \over 2 m} \right)^2 {q^2 \over 2 \pi^2} \left[ 
   -{4 \over 3} E_F p_F + \left( {q^2 \over 3} -2 M^2 \right)  \ln\left( 
   {E_F +p_F \over M} \right) +2 {E_F \over q} \left( {E_F^2 \over 3} +
   \varepsilon^2 \right) \ln A(q) \right.  \nonumber \\
   & & \left. \qquad \qquad \qquad
     +{8 \over 3}  {\varepsilon^3 \over q} \ln B(q) \right] \nonumber \\
   & & +\left( {f_\rho \over 2 m} \right) g_\rho {2 M q^2 \over \pi^2} \left[
    \ln\left( {E_F +p_F \over M} \right) -{E_F \over q} \ln A(q) -
   {\varepsilon \over q} \ln B(q) \right] 
\end{eqnarray}
for the transverse part.

\section*{Appendix B: Polarization, vacuum contribution}

\setcounter{equation}{0}

The vacuum contribution arises from the $-H(-p_0)$ term in the distribution 
function $f(p)$. This contribution is divergent and has to be renormalized.
After performing a dimensional regularization and subtracting the appropriate
counterterms, we obtained in section III a finite expression for the vacuum
polarization:

\beq
\Pi_{\rho\ vac}^{\mu\nu} &=& \left\{ {g_\rho^2 \over 3 \pi^2} 
\left[ \ln \left( {M \over m}  \right) +\theta(k^2,M^2) 
+{2 M^2 \over k^2} \left( \theta(k^2,M^2)-1 \right) \right] \right.
\nonumber \\
&& + \left( {f_\rho \over 2 m} \right)^2 {1 \over 6 \pi^2} 
\left[ 6 M^2  \ln \left( {M \over m}  \right) +8 M^2 \theta(k^2,M^2)
+k^2 \left( \ln \left( {M \over m}  \right) +\theta(k^2,M^2) \right) \right]
\nonumber \\
&& +{2 \over \pi^2} \left( {f_\rho \over 2 m} \right) g_\rho M
\left[ \ln \left( {M \over m}  \right) +\theta(k^2,M^2)  \right]
\nonumber \\
&& \left. + \alpha + \beta {(m-M) \over g_\sigma} + 
\gamma {(m-M)^2 \over g_\sigma^2} + \delta k^2 \right\} 
\left\{ k^2 g^{\mu\nu} -k^\mu k^\nu \right\}
\enq

with the function 
\begin{equation}
\theta (k^2, M^2)\ =\ \theta (y)\ \equiv\ y \int^\infty_0 {\frac{{dx}}{{\ %
\left[ (x^2\ +\ y). \sqrt{(x^2\ +\ 1)} \right] }}}
\end{equation}
where $y\ \equiv\ 1\ -\ k^2/(4M^2). $

The choice of conditions to be imposed on the polarization to determine the 
finite constants $\alpha$, $\beta$, $\gamma$, $\delta$ determine various
renormalization schemes.

In order to shorten the notations, we define 
\beq
\theta_\rho\ &=&\ \theta(\mu_\rho^2, m^2)\ , \qquad \theta_{\rho m}\ =\ 
\frac{\displaystyle \partial \theta (k^2, M^2)}{\displaystyle \partial M} 
\nonumber \\
\theta_{\rho k}\ &=&\ \frac{\displaystyle \partial \theta (k^2, M^2)}{%
\displaystyle \partial k^2}\ , \qquad \theta_{\rho mm}\ =\ \frac{%
\displaystyle \partial^2 \theta (k^2, M^2)}{\displaystyle \partial M^2} 
\enq
All these derivatives are to be evaluated in the point $k^2 = \mu_{\rho}^2$,
$M=m$.

We will note $\theta=\theta (k^2, M^2)$. Finally, we will work in units 
of the nucleon free mass, {\it i.e.} we set $m=1$ in all expressions 
given in this Appendix.

\vskip 0.5cm

$\underline{\mbox{\sf Scheme 1}}$

When the full set of conditions (\ref{ren1} -- \ref{ren4}) is applied to the 
total regularized vacuum polarization of the $\rho$ meson, the following 
expression is obtained 
\beq
\Pi_{\rho\ vac (1)}^{\mu\nu} &=& \left\{ {g_\rho^2 \over 3 \pi^2} \bigg[ 
\ln{M} + \theta + {2 M^2 \over k^2}(\theta-1) -{1 \over \mu_\rho^2}
\left(\, (2 + \mu_\rho^2)\, \theta_\rho -2  \right) \right. 
\nonumber \\
& & \qquad \quad +(1-M)\left( 1 + \theta_{\rho m} + 
{4 \over \mu_\rho^2} (\theta_\rho -1) +{2 \over \mu_\rho^2} 
\theta_{\rho m} \right) \nonumber \\
& & \qquad \quad -{1 \over 2} (1-M)^2 \left( -1  
+ \theta_{\rho m m} + {4 \over \mu_\rho^2}  (\theta_\rho-1) 
+{8  \over \mu_\rho^2} \theta_{\rho m} + {2  \over \mu_\rho^2}
 \theta_{\rho m m} \right) \nonumber \\
& &  \quad \qquad - (k^2 - \mu_\rho^2) \left( 1 +{2  \over \mu_\rho^2} 
\theta_{\rho k} -{2 \over \mu_\rho^4} (\theta_\rho -1)  
\right) \bigg] \nonumber \\
& & +\left( {f_\rho \over 2 m} \right)^2 {1 \over 6 \pi^2} \bigg[ 
6 M^2 \ln{M} + 8 M^2 \theta + k^2 ( \ln{M} + \theta) 
-(8  + \mu_\rho^2) \theta_\rho  \nonumber \\
& & \qquad \qquad \qquad \quad +(1-M)\left( 6\,  +16\,  \theta_\rho 
+ 8\,  \theta_{\rho m} +
\mu_\rho^2 (1  + \theta_{\rho m} ) \right) \nonumber \\
& & \qquad \qquad \qquad \quad -{1 \over 2} (1-M)^2 
\left(18 +16\, \theta_\rho + 32\,  \theta_{\rho m} + 8  \theta_{\rho m m} 
+ \mu_{\rho}^2 (-1 + \theta_{\rho m m}) \right) \nonumber \\
& & \qquad \qquad \qquad \quad  -(k^2 - \mu_\rho^2) \left(\, 
( 8  + \mu_\rho^2)\, \theta_{\rho k} + \theta_\rho \right) \bigg] \nonumber \\
& & +{2 \over \pi^2} \left( {f_\rho \over 2 m} \right) g_\rho \bigg[
M ( \ln{M } + \theta) -  \theta_\rho + (1-M)\left( \theta_\rho
+ 1 + \theta_{\rho m} \right)  \nonumber \\
& & \left. \qquad \qquad \qquad \quad -{1 \over 2} (1-M)^2 \left(  1  
+ 2\, \theta_{\rho m}   + \theta_{\rho m m}  \right) 
- (k^2 - \mu_\rho^2)\  \theta_{\rho k} \bigg]  \right\} \left\{
k^2 g^{\mu\nu} - k^{\mu} k^\nu \right\}
\enq

\vskip 0.5cm

$\underline{\mbox{\sf Scheme 2}}$

\vskip 0.5cm

The former scheme has the drawback that we do not recover the expression
of the vacuum polarization of the $\omega$ meson when we take the limit
$f_\rho \rightarrow 0$ and replace $g_\rho$ by $g_\omega$. The reason is 
that the structure of the infinities is not the same for the different 
couplings. By renormalizing all 
contributions together we are in fact introducing spurious counterterms 
where they need not be. For example, in the $(f_\rho/2 m)^2$ contribution 
we have both $M^2 /\epsilon$ and $k^2 /\epsilon$, needing all sets of 
counterterms to cancel them, whereas in the $g_\rho^2$ contribution we have 
a $1/\epsilon$ which requires only the $A\ {\vec{R}}^{\mu \nu }\,\cdot 
\,{\vec{R}}_{\mu \nu }$ counterterm.
This can easily be cured by splitting the counterterms as follows
$A=A_{gg}+A_{fg}+A_{ff}$, $B=B_{fg}+B_{ff}$, $C=C_{ff}$, $D=D_{ff}$, 
(where $A_{fg}$, $B_{fg}$ mean terms proportional to 
$g_\rho\, f_\rho$, and so on) and renormalized separately each contribution. 
We obtain

\beq
\Pi_{\rho\ vac (2)}^{\mu\nu} &=& \left\{ {g_\rho^2 \over 3 \pi^2} \bigg[ 
\ln{M } + \theta + {2 M^2 \over k^2}(\theta-1) -{1 \over \mu_\rho^2}
\left(\, (2 + \mu_\rho^2)\, \theta_\rho -2 \right)  \bigg] 
\right. \nonumber \\
& & +\left( {f_\rho \over 2 m} \right)^2 {1 \over 6 \pi^2} \bigg[ 
6 M^2 \ln{M } + 8 M^2 \theta + k^2 ( \ln{M } + \theta) 
-(8 + \mu_\rho^2)\, \theta_\rho  \nonumber \\
& & \qquad \qquad \qquad \quad +(1-M)\left( 6 +16\, \theta_\rho 
+ 8\, \theta_{\rho m} + \mu_\rho^2 (1  + \theta_{\rho m} ) \right) 
\nonumber \\
& & \qquad \qquad \qquad \quad -{1 \over 2} (1-M)^2 
\left(18 +16\, \theta_\rho + 32\, \theta_{\rho m} + 8\, \theta_{\rho m m} 
+ \mu_{\rho}^2 (-1 + \theta_{\rho m m}) \right) \nonumber \\
& &  \qquad \qquad \qquad \quad  -(k^2 - \mu_\rho^2) 
\left(\, ( 8 + \mu_\rho^2)\, 
\theta_{\rho k} + \theta_\rho \right) \bigg] \nonumber \\
& &  \left. +{2 \over \pi^2} \left( {f_\rho \over 2 m} \right) g_\rho \bigg[
M ( \ln{M } + \theta) -  \theta_\rho + (1-M)\left( \theta_\rho
+ 1 + \theta_{\rho m} \right)  \bigg]  \right\} \left\{
k^2 g^{\mu\nu} - k^{\mu} k^\nu \right\}
\enq

\vskip 0.5cm

$\underline{\mbox{\sf Scheme 3}}$

We observe that there appear infinities which can be written as depending
on $M$ (rather than on $\sigma$), as for example $M^2/\epsilon$. 
After absorbing the infinities, the finite part of the counterterms is 
$a_{ff} + b_{ff} \sigma + c_{ff} \sigma^2$. The constants $a_{ff}$, $b_{ff}$,
$c_{ff}$ which are calculated by imposing the conditions (\ref{ren1}) and 
(\ref{ren3},\ref{ren4}) cannot be cast any more in a form proportional
to $M^2$. One would like to require that the $M^2$ structure be preserved 
by dropping conditions (\ref{ren3},\ref{ren4}). Proceeding in this way, 
we obtain

\beq
\Pi_{\rho\ vac (3)}^{\mu\nu} &=& \left\{ {g_\rho^2 \over 3 \pi^2} \bigg[ 
\ln M + \theta + {2 M^2 \over k^2}(\theta-1) -{1 \over \mu_\rho^2}
\left(\, (2 + \mu_\rho^2)\, \theta_\rho -2  \right)  \bigg] 
\right. \nonumber \\
& & +\left( {f_\rho \over 2 m} \right)^2 {1 \over 6 \pi^2} \bigg[ 
6 M^2 \ln M  + 8 M^2 \theta + k^2 ( \ln M  + \theta) -(8+\mu_\rho^2)\, 
\theta_\rho   \nonumber \\
& & \qquad \qquad \qquad \quad 
-(k^2 - \mu_\rho^2) \left(\, (8 + \mu_\rho^2)\, \theta_{\rho k} 
+ \theta_\rho  \right)
+(M^2-1) \left( \mu_\rho^2 (8 + \mu_\rho^2)\, \theta_{\rho k} 
-8  \theta_\rho  \right)  \bigg] \nonumber \\
& &  \left. +{2 \over \pi^2} \left( {f_\rho \over 2 m} \right) g_\rho \bigg[
M ( \ln M  + \theta  -  \theta_\rho )  \bigg]  \right\} \left\{
k^2 g^{\mu\nu} - k^{\mu} k^\nu \right\}
\enq

\vskip 0.5cm

$\underline{\mbox{\sf Scheme 4}}$

We show in this section and the next how the results of Shiomi and Hatsuda
\cite{[SH94]} and Sarkar {\it et al.} \cite{Sarkar98} can be recovered.
We first observe that the regularized vacuum polarization can be written
as follows:

\beq 
\Pi_{\rho\ vac}^{\mu\nu} =   \left\{ g_\rho^2 {\cal A} +
({ f_\rho \over 2 m} )^2 \left[ {k^2 \over 2} {\cal A} 
+ {M^2 \over 2} {\cal B} \right] +  g_\rho ( {f_\rho \over 2 m} ) M {\cal B} 
\right\} \left( g^{\mu\nu} -{k^\mu k^\nu \over k^2} \right)
\label{regstruct}
\enq
with 
\beq
{\cal A} &=& {2 \over (2 \pi)^3} {1\over k^2} 
\left[ -16 {\cal I}_1 /3 -2/3 k^2 (2 M^2 +k^2) {\cal I}_2 \right] \nonumber \\
{\cal B} &=& {2 \over (2 \pi)^3} \left[ -4 k^2 {\cal I}_2 \right] \nonumber
\enq

If we now impose that this structure be kept after the renormalization,
with ${\cal A}$ and ${\cal B}$ substituted by finite $\widetilde{\cal A}$ 
and $\widetilde{\cal B}$ after extracting the infinities, we can determine the
constants by imposing that $\widetilde{\cal A}$ and $\widetilde{\cal B}$ 
vanish on the mass shell. By so doing we obtain

\beq
\widetilde{\cal A} &=& {1 \over 3 \pi^2} \left[ \ln(M)+\theta + {2 M^2 \over k^2}
(\theta -1) -{1 \over \mu_\rho^2} \left( (2+\mu_\rho^2) \theta_\rho -2 \right) 
\right] \nonumber \\
\widetilde{\cal B} &=& {2 k^2 \over \pi^2} \left[ \ln(M)+\theta - \theta_\rho 
\right] \nonumber
\enq
and
\beq
\Pi_{\rho\ vac (4)}^{\mu\nu}&=& \left\{ {g_\rho^2 \over 3 \pi^2} \left[ 
\ln M + \theta + {2 M^2 \over k^2}(\theta-1) -{1 \over \mu_\rho^2}
\left( (2 + \mu_\rho^2) \theta_\rho -2  \right)  \right] 
\right. \nonumber \\
& & +\left( {f_\rho \over 2 m} \right)^2 {1 \over 6 \pi^2} \left[ 
6 M^2 \ln M  + 8 M^2 \theta + k^2 ( \ln M  + \theta) -2 M^2 
-6 M^2 \theta_\rho \right. \nonumber \\
& & \qquad \quad \left.
-{k^2 \over \mu_\rho^2} \left( (2 + \mu_\rho^2) \theta_\rho -2  \right)
 \right] \nonumber \\
& &  \left. +{2 \over \pi^2} \left( {f_\rho \over 2 m} \right) g_\rho \left[
M ( \ln M  + \theta  -  \theta_\rho )  \right]  \right\} \left\{
k^2 g^{\mu\nu} - k^{\mu} k^\nu \right\}
\label{renormsarkar}
\enq

We now use the fact that our function $\theta$ is related to the 
integrals appearing in the expressions of Shiomi and Hatsuda and 
Sarkar {\it et al.}.

\beq
\int_0^1 dx \ln[M^2 -k^2 x (1-x)] &=& -2 +2 \ln(M) +2 \theta \nonumber \\
\int_0^1 dx x (1-x) \ln[M^2 -k^2 x (1-x)] &=& -{5 \over 8} -{2 \over 3} 
{M^2 \over k^2} +{1 \over 3} \ln(M) -{1 \over 3} {(2 M^2 + k^2) \over k^2} \theta
\nonumber
\enq
With these relations, it is straightforward to check that Eq. (\ref{renormsarkar})
coincides with the expression of Sarkar {\it et al.}

\vskip 0.5cm

$\underline{\mbox{\sf Scheme 5}}$

The expression of Shiomi and Hatsuda can be recovered by the following 
recipe: First, one should subtract from the expressions of the 
functions ${\cal A}$, ${\cal B}$ from the preceding paragraph their 
values in the true vacuum $M=m$, and then, replace them in the 
expression\footnote{We note that this is not simply
the same as subtracting the true vacuum from the total polarization,
since the effective mass appears outside of $\widetilde{\cal A}$, 
$\widetilde{\cal B}$ in Eq. (\ref{regstruct})} Eq. (\ref{regstruct}). 
In terms of our $\theta$ functions, one has
\beq
\Pi_{\rho\ vac (5)}^{\mu\nu} &=& \Pi_{\rho\ vac 
(S\mbox{\rm\tiny \&}H)}^{\mu\nu}=
\left\{ {g_\rho^2 \over 3 \pi^2} \left[ 
\ln M + \theta + {2 M^2 \over k^2}(\theta-1) -{2 \over k^2} 
\left(\theta_0 -1 \right) -  \theta_0 \right] 
\right. \nonumber \\
& & +\left( {f_\rho \over 2 m} \right)^2 {1 \over 6 \pi^2} \left[ 
6 M^2 \ln M  + 8 M^2 \theta + k^2 ( \ln M  + \theta) -2 (M^2 -1) 
-(6 M^2 +k^2 +2) \theta_0 \right] \nonumber \\
& &  \left. +{2 M \over \pi^2} \left( {f_\rho \over 2 m} \right) 
g_\rho \left[ ( \ln M  + \theta  -  \theta_0 )  \right]  \right\} 
\left\{k^2 g^{\mu\nu} - k^{\mu} k^\nu \right\}
\enq

\vskip 0.5cm

$\underline{\mbox{\sf Scheme 6}}$

Another standard renormalization scheme is that used by Kurasawa 
and Suzuki \cite{[KS88]} in the case of the $\sigma$ and $\omega$ 
mesons. In this scheme, the conditions Eqs. (\ref{ren1},\ref{ren2}) 
are applied at the physical mass $k^2=\mu_\rho^2$ whereas conditions 
(\ref{ren3},\ref{ren4}) are taken at $k^2=0$.
The following expression is obtained 
\beq
\Pi_{\rho\ vac (6)}^{\mu\nu} &=& \left\{ {g_\rho^2 \over 3 \pi^2} \bigg[ 
\ln{M} + \theta + {2 M^2 \over k^2}(\theta-1) -{1 \over \mu_\rho^2}
\left(\, (2 + \mu_\rho^2)\, \theta_\rho -2  \right)  \bigg] 
\right. \nonumber \\
& & +\left( {f_\rho \over 2 m} \right)^2 {1 \over 6 \pi^2} \bigg[ 
6 M^2 \ln{M} + 8 M^2 \theta + k^2 ( \ln{M} + \theta) 
-(8  + \mu_\rho^2) \theta_\rho  \nonumber \\
& & \qquad \qquad \qquad \quad  -(k^2 - \mu_\rho^2) \left(\, 
( 8  + \mu_\rho^2)\, \theta_{\rho k} + \theta_\rho \right) 
 +5 +12 M -17 M^2 \bigg] \nonumber \\
& & \left. +{2 \over \pi^2} \left( {f_\rho \over 2 m} \right) g_\rho \bigg[
M ( \ln{M } + \theta) -  \theta_\rho + 2 (1-M) \bigg]  \right\} \left\{
k^2 g^{\mu\nu} - k^{\mu} k^\nu \right\}
\enq

\section*{Appendix C: Interaction potential in spatial coordinates}

\setcounter{equation}{0}

In this appendix, we give explicit expressions for the different meson
contributions to the several pieces of the potential, as they appear in Eq.
(\ref{V(r)}) : central, spin-orbit, spin-spin, tensor, and the nonlocal
potential $V_{NL}(r)$.

A subscript will label these components, while the superscript indicates the
meson which gives this contribution.

These expressions are valid when the rest frame of the background fluid
coincides with the center of mass of the interacting particles.

\vskip 0.5cm 

$\underline{\mbox{\sf  Rho meson}}$

\vskip 0.5cm 

\beq
V_c^{(\rho)}(r) &=& \quad {1 \over 2 \pi^2 r} \int_0^\infty dq\, q \sin(qr)\, 
{\cal F}_{\rho}^2 \left\{ -g_\rho^2\ G_{\rho\, L} 
\left( 1 -{q^2 \over 4 M} \right) + g_\rho \left( {f_\rho \over 2 m} \right)
\ G_{\rho\, L} \left( {q^2 \over M}\right)  \right. \nonumber \\
& & \left. \qquad \qquad \qquad \qquad \qquad \quad
+ g_\rho^2\  G_{\rho\, T} \, \left( 
{q^2 \over 4 M^2} \right) \right\} \\
& & \nonumber \\
V_{SS}^{(\rho)}(r) &=& \quad \left[ g_{\rho} + 2M\,  \left( {f_\rho \over 2 m} 
\right) \right]^{2} \ {1 \over 12 \pi^2  M^2 \ r} \  \int_0^\infty dq \,
q^3 \, \sin(qr) \, {\cal F}_{\rho}^2 \  \ G_{\rho\, T} \\
& & \nonumber \\
V_T^{(\rho)}(r)&=& \quad \left[ g_{\rho} + 2M\,  \left( {f_\rho \over 2 m} 
\right) \right]^{2} \ {1 \over 24 \pi^2  M^2  r} \  \int_0^\infty dq \,
 \sin(qr) \, {\cal F}_{\rho}^2 \left[ {3 q \over r^2} -q^3 \right]\, 
G_{\rho\, T} \nonumber \\
& & - \left[ g_{\rho} + 2M\,  \left( {f_\rho \over 2 m} 
\right) \right]^{2} \ {1 \over 8 \pi^2  M^2  r^2} \  \int_0^\infty dq \,
 q^2\, \cos(qr) \, {\cal F}_{\rho}^2 \, G_{\rho\, T} \\
& & \nonumber \\
V_{LS}^{(\rho)}(r) &=&  \quad {1 \over 2 \pi^2 M^2 r^3 } \int_0^\infty 
dq\, q\, \sin(qr)\, {\cal F}_{\rho}^2 \left\{ g_\rho^2 \, 
\left[ {G_{\rho\, L} \over 2} + G_{\rho\, T} \right] +2 M\, g_\rho 
\left( {f_\rho \over 2 m} \right) \left[ G_{\rho\, L} + G_{\rho\, T} \right]
\right\} \nonumber \\
& & - {1 \over 2 \pi^2 M^2 r^2 } \int_0^\infty 
dq\, q^2 \cos(qr)\, {\cal F}_{\rho}^2 \left\{ g_\rho^2 \, 
\left[ {G_{\rho\, L} \over 2} + G_{\rho\, T} \right] +2 M\, g_\rho 
\left( {f_\rho \over 2 m} \right) \left[ G_{\rho\, L} + G_{\rho\, T} \right]
\right\} \\
& & \nonumber \\
V_{NL}^{(\rho)}(r)\,&=&- \frac{\displaystyle g_{\rho}^2}{\displaystyle 2
\,\pi^2 \, M^2 \,r} \,  \int^{\infty}_{0} dq \,q \,sin(qr) \, {\cal F}_{\rho}^2
\left[ {G_{\rho\, L} \over 2} +G_{\rho\, T} \right]  
\enq

with the following notations

\begin{eqnarray}
G_{\rho L} =-G^{00}_{\rho}(q) = \frac{\displaystyle -1}{\displaystyle 
q^2+\mu_\rho^2-\Pi_\rho^{00}(0,q)} \qquad ,\qquad 
G_{\rho T}=G^{11}_{\rho}(q) = \frac{\displaystyle -1}{%
\displaystyle q^2+\mu_\rho ^2+ \Pi_\rho^{11}(0,q)}  \nonumber
\end{eqnarray}

\vskip 0.5cm 

$\underline{\mbox{\sf  Sigma and omega mesons}}$

\vskip 0.5cm 

The mixed $\sigma$-$\omega$ sector gives

\beq
V_c^{(\sigma \omega)}(r) &=& \quad {1 \over 2 \pi^2 r} \int_0^\infty\, dq\, q\,
\sin(q r) \left\{ g_\sigma^2 {\cal F}_\sigma^2\, G_\sigma 
\left( 1 +{q^2 \over 4 M^2} \right) - g_\omega^2 {\cal F}_\omega^2\,
 G_{\omega\, L} \left( 1-{q^2 \over 4 M^2} \right) \right. \nonumber \\
& & \left. \qquad \qquad \qquad \qquad
+ g_\omega^2\, {\cal F}_\omega^2 \, G_{\omega\, T} \left( {q^2 \over 4 M^2} 
\right) -2 g_\sigma g_\omega\, {\cal F}_\omega {\cal F}_\sigma\, 
G_{\sigma\omega} \right\} \\
& & \nonumber \\
V_{LS}^{(\sigma \omega)}(r) &=&  \quad {1 \over 2 \pi^2 M^2 r^3 } 
\int_0^\infty dq\, q\, \sin(qr)\, \left\{ g_\sigma^2\, {\cal F}_{\sigma}^2\,
{G_\sigma\over 2} + g_\omega^2 \, {\cal F}_{\omega}^2 
\left[ {G_{\omega\, L} \over 2} + G_{\omega\, T} \right] \right\} \nonumber \\
& & - {1 \over 2 \pi^2 M^2 r^2 } \int_0^\infty dq\, q^2\, \cos(qr)\, 
\left\{ g_\sigma^2\, {\cal F}_{\sigma}^2\, {G_\sigma\over 2} 
+ g_\omega^2 \, {\cal F}_{\omega}^2  \left[ {G_{\omega\, L} \over 2} 
+ G_{\omega\, T} \right] \right\} \\
& & \nonumber \\
V_{NL}^{(\sigma \omega)}(r)&=& -{1 \over 2 \pi^2 M^2 r} \int_0^\infty dq\, 
q\, \sin(qr)\, \left\{ g_\sigma^2\, {\cal F}_{\sigma}^2\, {G_\sigma \over 2}
+ g_\omega^2 \, {\cal F}_{\omega}^2 \left[ {G_{\omega\, L} \over 2} 
+ G_{\omega\, T} \right] \right\} \\
& & \nonumber \\
V_{SS}^{(\sigma \omega)}(r)&=&\quad {g_\omega^2 \over 12 \pi^2 M^2 r} 
\int_0^\infty dq\, q^3\, \sin(q r)\, {\cal F}_\omega^2\, G_{\omega\, T} \\
& & \nonumber \\
V_{T}^{(\sigma \omega)}(r) & =  &  \quad  {g_\omega^2 \over 24 \pi^2  M^2  r} 
\ \int_0^\infty dq \, \sin(qr) \, {\cal F}_{\omega}^2 \left[ {3 q \over r^2} 
-q^3 \right]\, G_{\omega\, T} \nonumber \\
& & -  {g_\omega^2 \over 8 \pi^2  M^2  r^2} \  \int_0^\infty dq \,
 q^2\, \cos(qr) \, {\cal F}_{\omega}^2 \, G_{\omega\, T} 
\enq

In the above formulae, one has

\beq
G_{\sigma} &=&  -\, { q^2+\mu_{\omega}^2-\Pi^{00}_{\omega}(0,q) \over
\left( q^2+\mu_{\sigma}^2+\Pi_{\sigma}(0,q)     \right)
\left( q^2+\mu_{\omega}^2-\Pi^{00}_{\omega}(0,q)\right) 
+  \left( \Pi_{\omega\sigma}^{0}(0,q) \right)^2 } \nonumber \\
G_{\sigma\omega} &=& -\, { \Pi_{\omega\sigma}^{0}(0,q) \over 
\left( q^2+\mu_{\sigma}^2+\Pi_{\sigma}(0,q)     \right)
\left( q^2+\mu_{\omega}^2-\Pi^{00}_{\omega}(0,q)\right) 
+  \left( \Pi_{\omega\sigma}^{0}(0,q) \right)^2 } \nonumber \\
G_{\omega\, L} &=& -\, {q^{2}+\mu_{\sigma}^{2}+{\Pi}_{\sigma}(0,q) \over
\left( q^2+\mu_{\sigma}^2+\Pi_{\sigma}(0,q)     \right)
\left( q^2+\mu_{\omega}^2-\Pi^{00}_{\omega}(0,q)\right) 
+  \left( \Pi_{\omega\sigma}^{0}(0,q) \right)^2 } \nonumber \\
G_{\omega\, T} &=& -\, { 1 \over q^2+\mu_{\omega}^2+\Pi^{11}_{\omega}(0,q)}
\nonumber 
\enq

\vskip 0.5cm 

$\underline{\mbox{\sf  Pi meson}}$

\vskip 0.5cm 

Finally, the one-pion exchange only contributes to the spin-spin and tensor
components :

\beq
V_{SS}^{\, \pi}(r)\,&=&\, \frac{\displaystyle g_{\pi}^{2}}{\displaystyle 24
\pi^2 M^2 r} \ {\displaystyle \int}^{\infty}_{0} dq 
\,q^{3} \, \sin(qr) \, {\cal F}_{\pi}^2\, G_{\pi} \\
& & \nonumber \\
V_{T}^{\, \pi}(r) &=&\  \frac{\displaystyle -g_{\pi}^{2}}{\displaystyle 24
\pi^2 M^2 r} \, {\displaystyle \int}^{\infty}_{0}dq  \, 
\left[ \frac{\displaystyle 3 q}{\displaystyle r^{2}}- q^{3} \right]
\, \sin(qr) \,{\cal F}_{\pi}^2\, G_{\pi}  \nonumber \\
& & + \frac{\displaystyle g_{\pi}^{2}}{\displaystyle 8 \pi^2 M^2 r^{2}} {%
\displaystyle \int}^{\infty}_{0} dq \, q^{2}\, \cos(qr) \,{\cal F}_{\pi}^2\,
G_{\pi}
\enq

where :

\beq
G_\pi(q) =  {\frac{{-1}}{{q^2 \ + \ \mu_{\pi}^2 \ +\ {\Pi}_{\pi}
(0,q) }}} \nonumber
\enq

\newpage

\mafigura{9 cm}{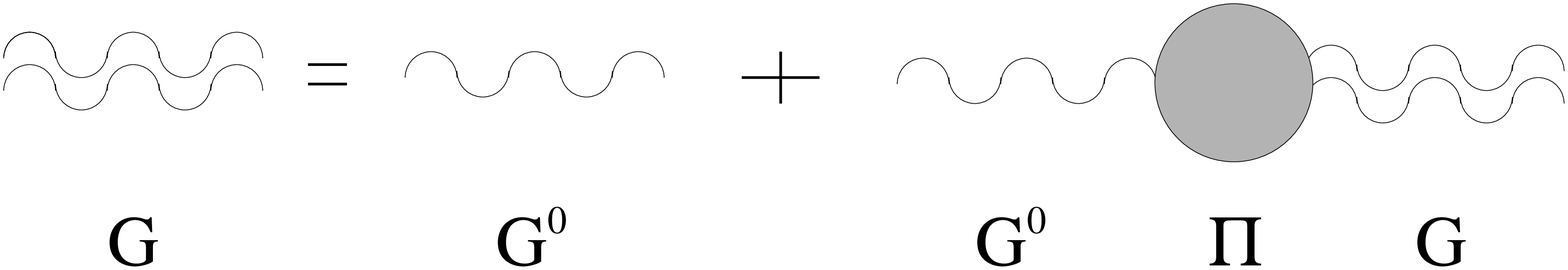}{Schematic representation of the Schwinger-Dyson 
equation for the meson propagators.}{Fig. 1}

\mafigura{7cm}{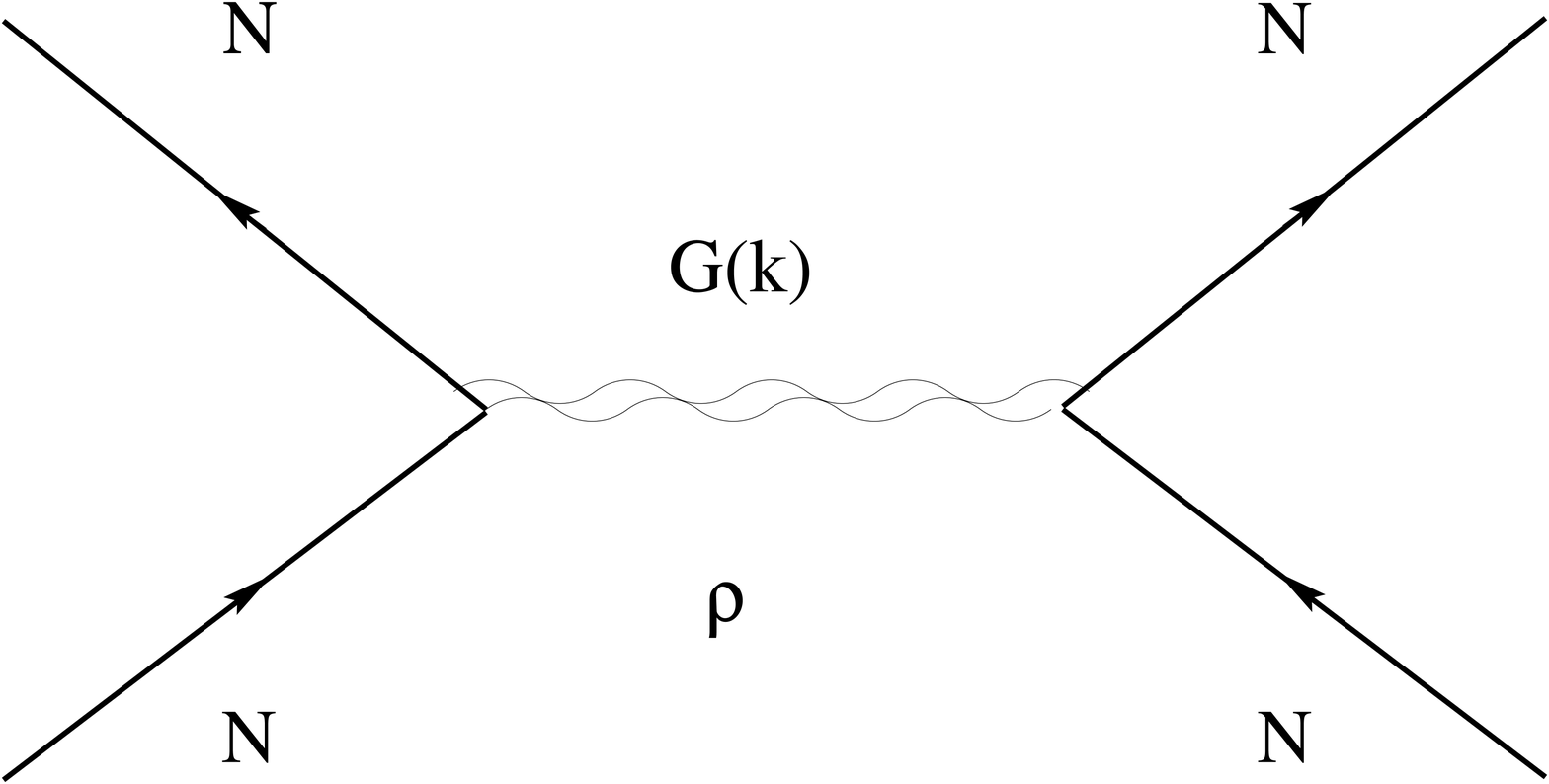}{Diagram for one-rho meson exchange. The double
line represents the in-medium rho meson propagator.}{Fig. 2}

\mafigura{14cm}{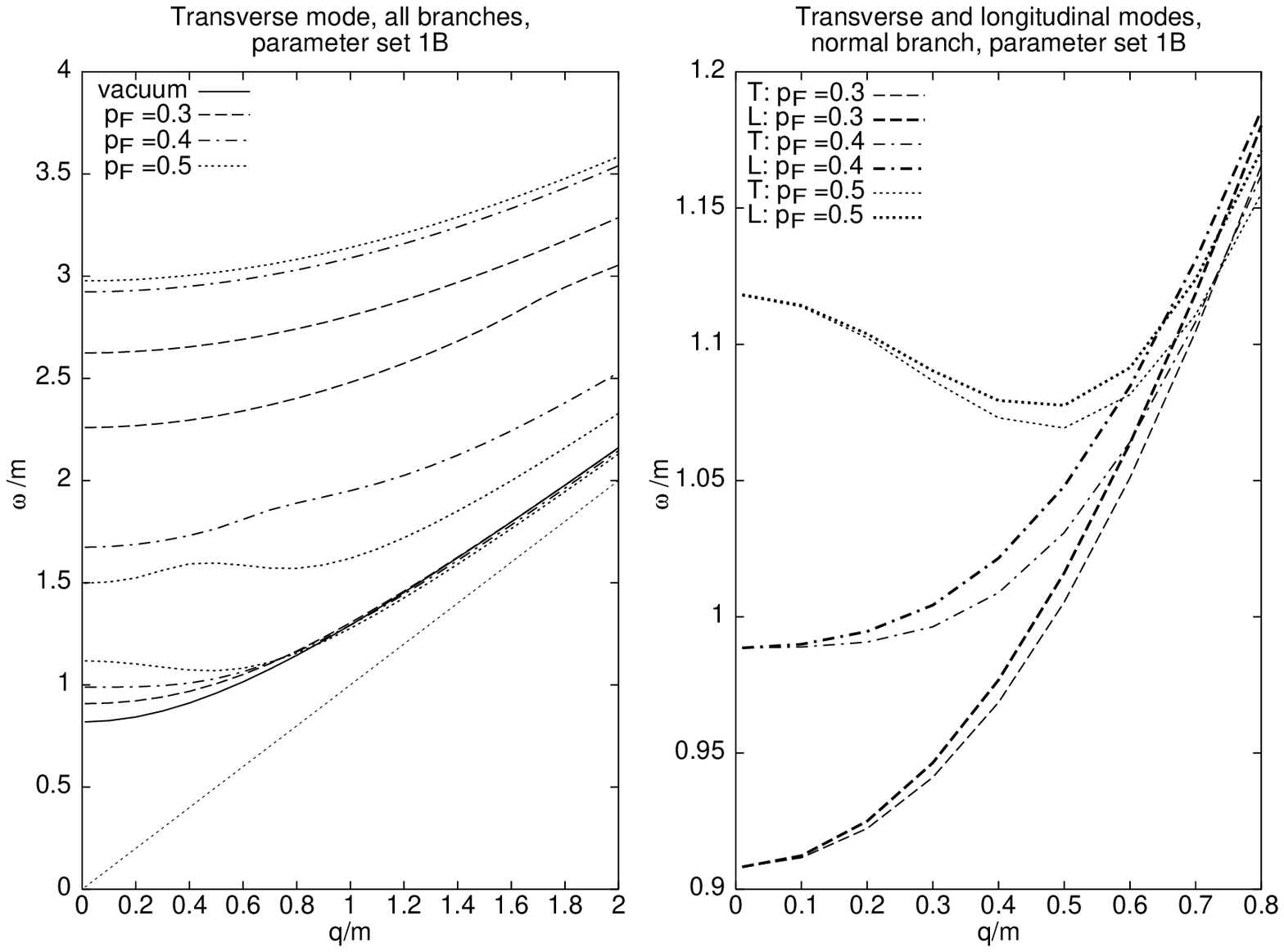}{Rho-meson dispersion relations for parameter set 1B. 
The solid line correspond to vacuum density, and the dashed, dot-dashed and 
dotted lines to finite density with a value of Fermi momentum $p_F/m$=0.3, 0.4, 0.5
respectively. The left panel shows transverse modes. There are normal and 
heavy meson branches. The right panel compares longitudinal (thick lines)
to transverse modes (thin lines) for the normal branch. We use the same 
conventions as in the left panel to represent various densities. 
Magnitudes are given in units of the free nucleon mass $m$.}{Fig. 3}

\mafigura{15cm}{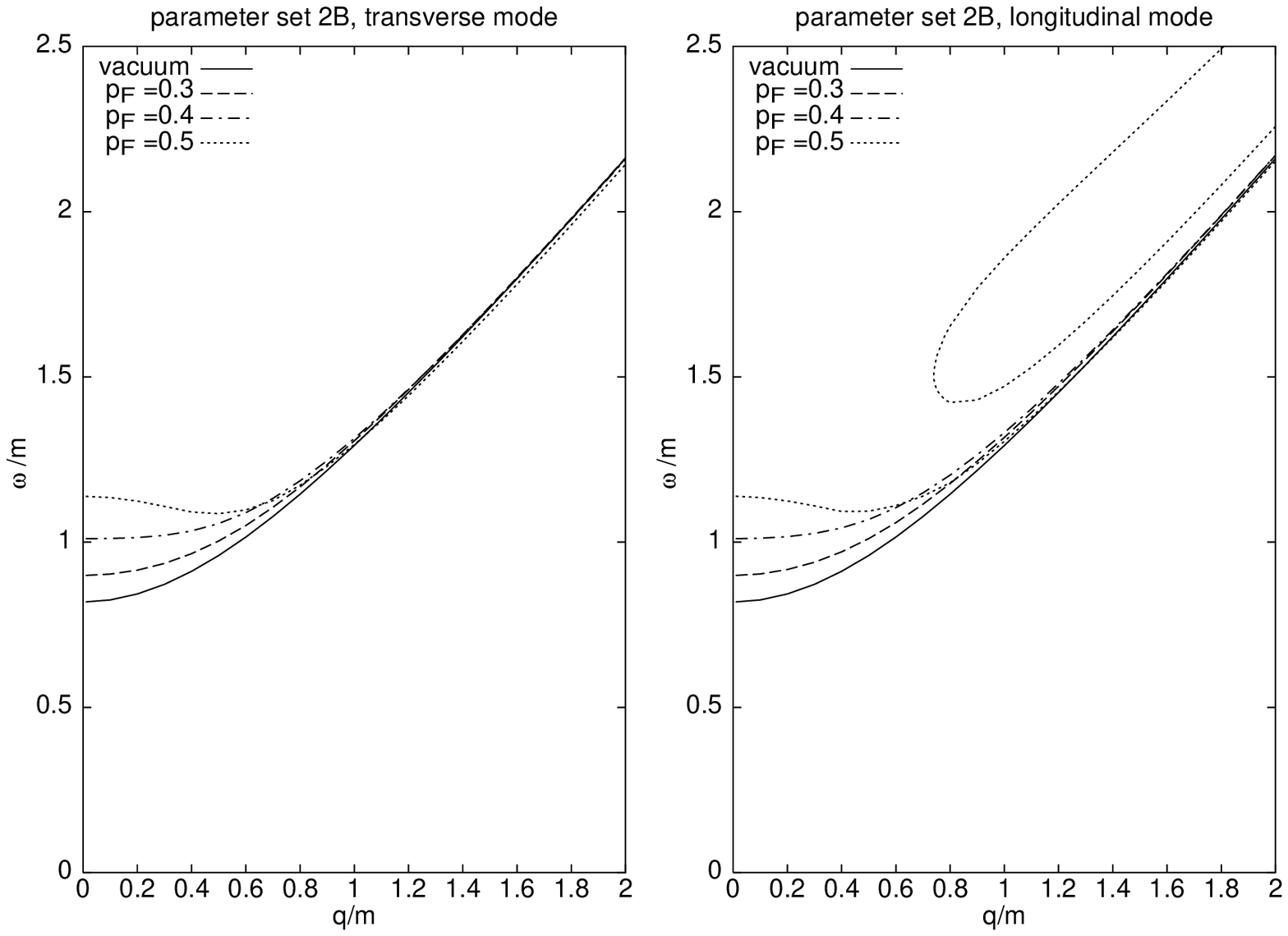}{Rho-meson dispersion relations for parameter set 2B.
The left panel shows the transverse branch at various density with
the same conventions as in Fig. 3. Only the normal branch remains in this
case. The longitudinal branches are represented in the right panel. Besides the
normal branch, we have a remnant of the heavy meson modes at high density.}{Fig. 4}

\mafigura{8cm}{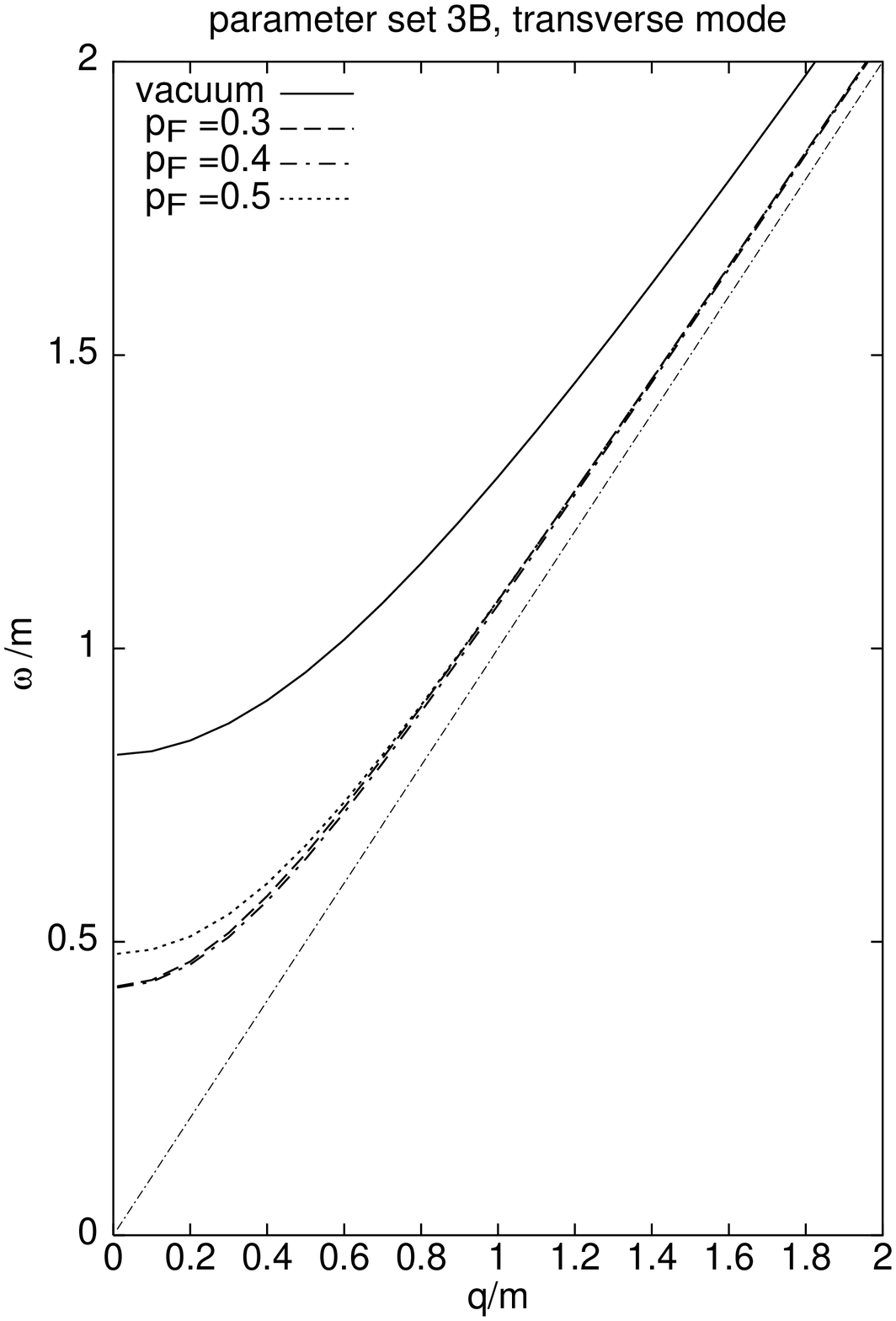}{Rho-meson dispersion relation for parameter set 3B.
Only the transverse branch is shown. The longitudinal branch almost coincides with
the transverse one so that it could not be distinguished by eye from the former.
Only normal branches are present in this case}{Fig. 5}

\mafigura{10cm}{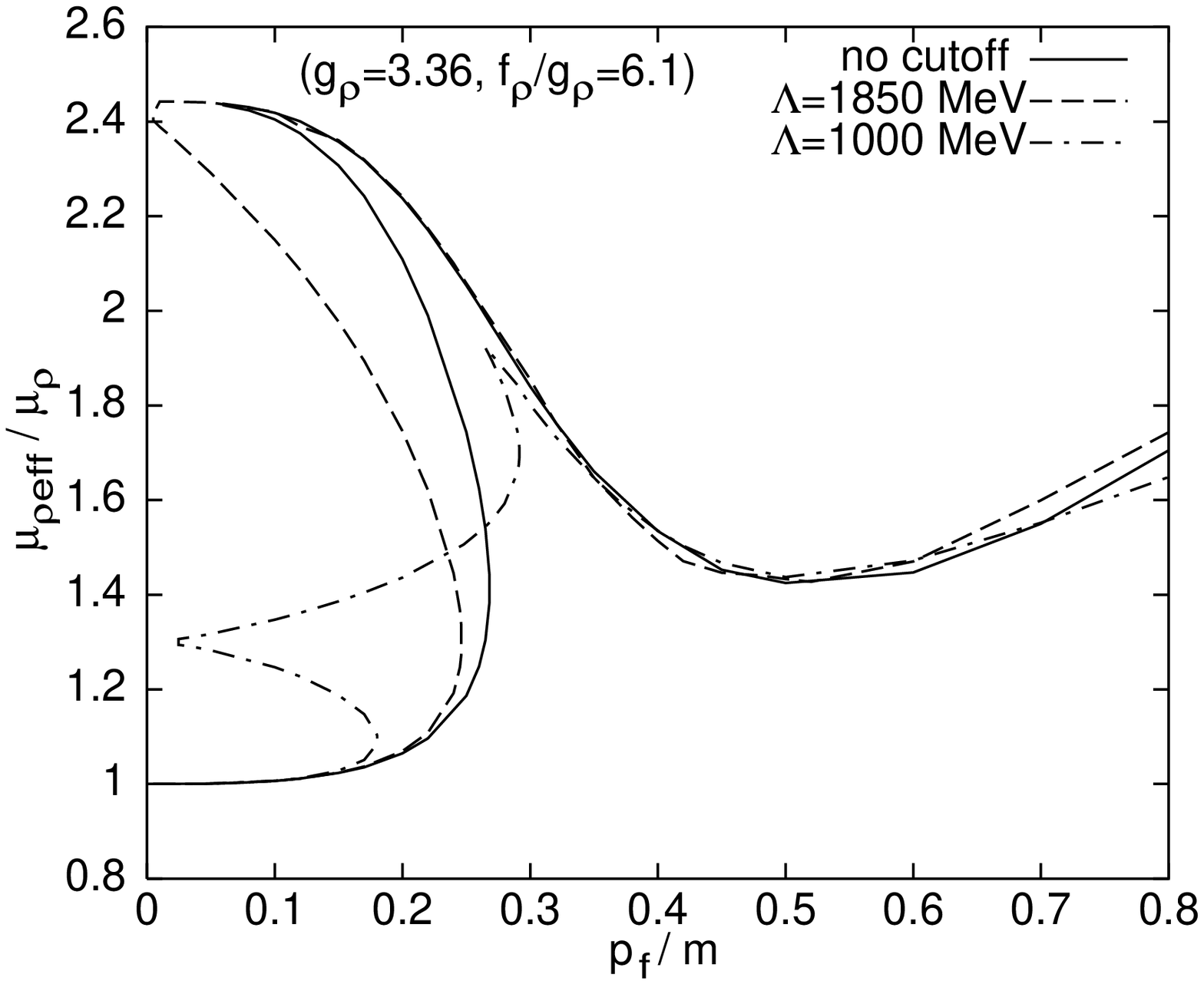}{In-medium $\rho$-meson mass as a function of the 
nuclear Fermi momentum, when the vacuum term is discarded. For small densities,
it would appear that the effective rho mass increases; however neglecting
of vacuum fluctuations is at the origin of an unpleasant cusp, which cannot be 
removed even by very low values of the cutoff parameter in the form factor}{Fig. 6}

\mafigura{10cm}{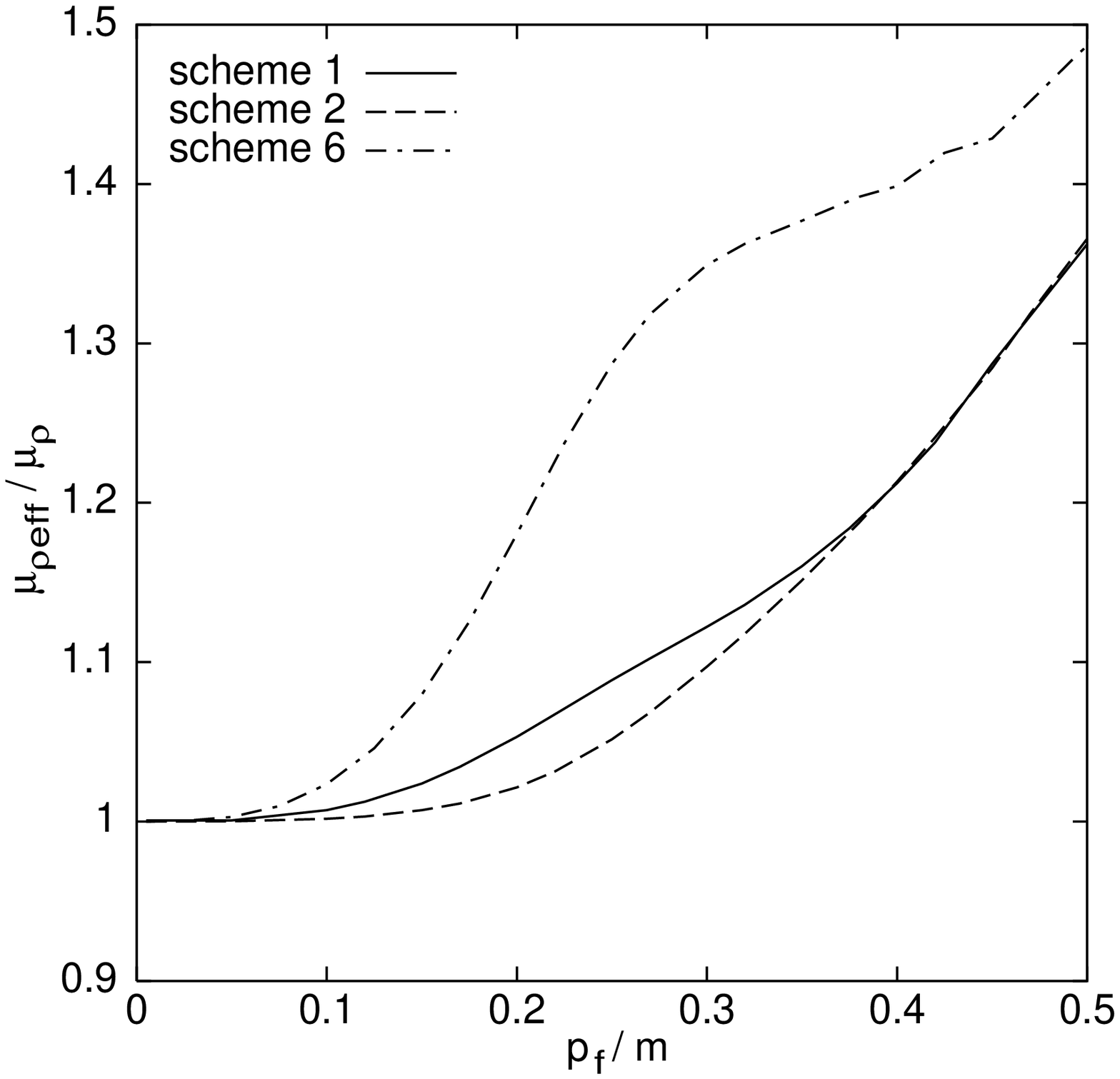}{In-medium $\rho$-meson mass as a function of the 
nuclear Fermi momentum. Vacuum polarization is included according to the first 
class of renormalization schemes 1, 2 and 6 (= ``Kurasawa-Suzuki'').}{Fig. 7}

\mafigura{10cm}{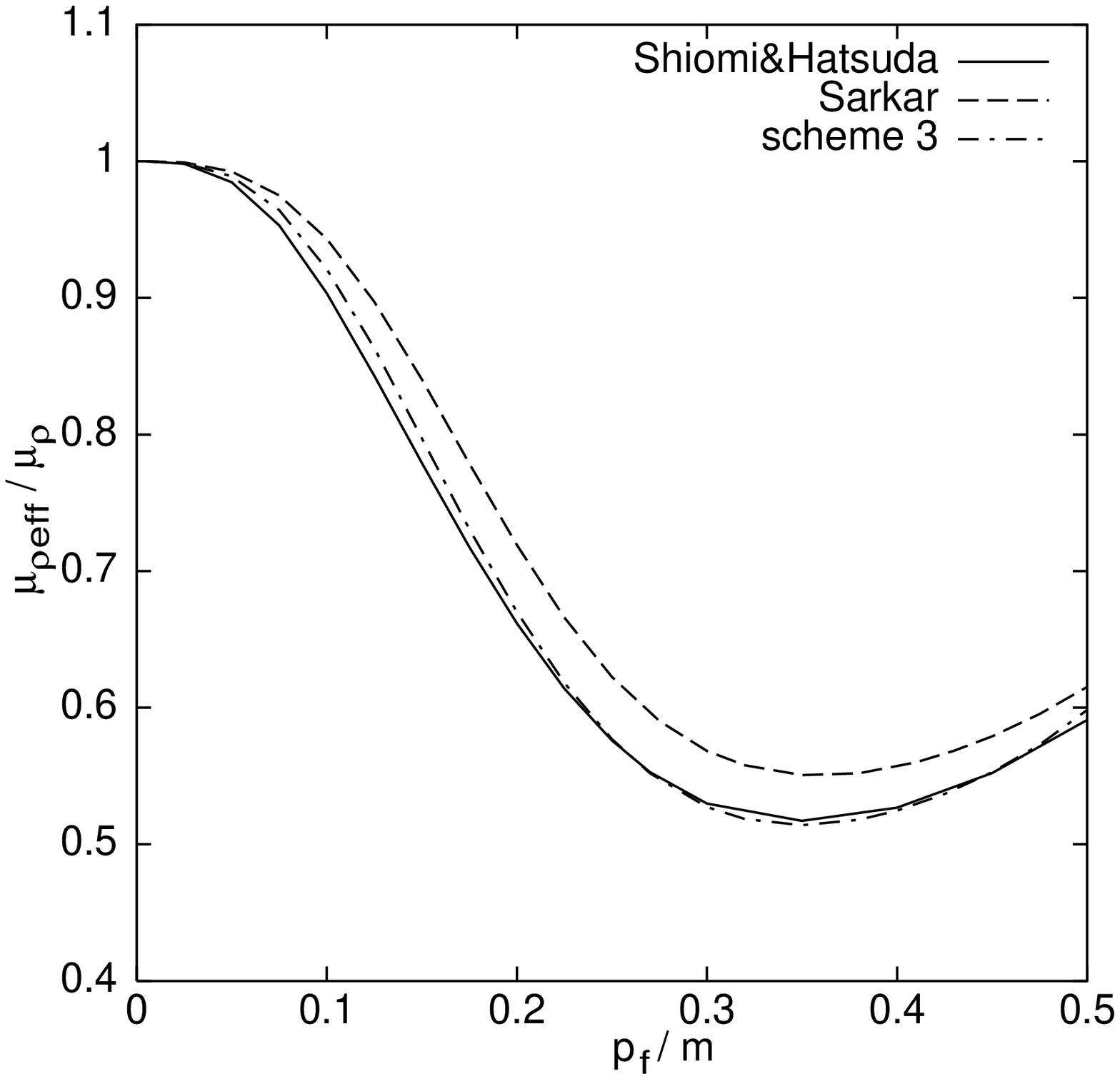}{In-medium $\rho$-meson mass as a function of the 
nuclear Fermi momentum. Vacuum polarization is included according to the 
second  class of renormalization schemes 3, 4 (= `` Sarkar'') and 5 
(= ``Shiomi-Hatsuda'').}{Fig. 8}

\mafigura{16cm}{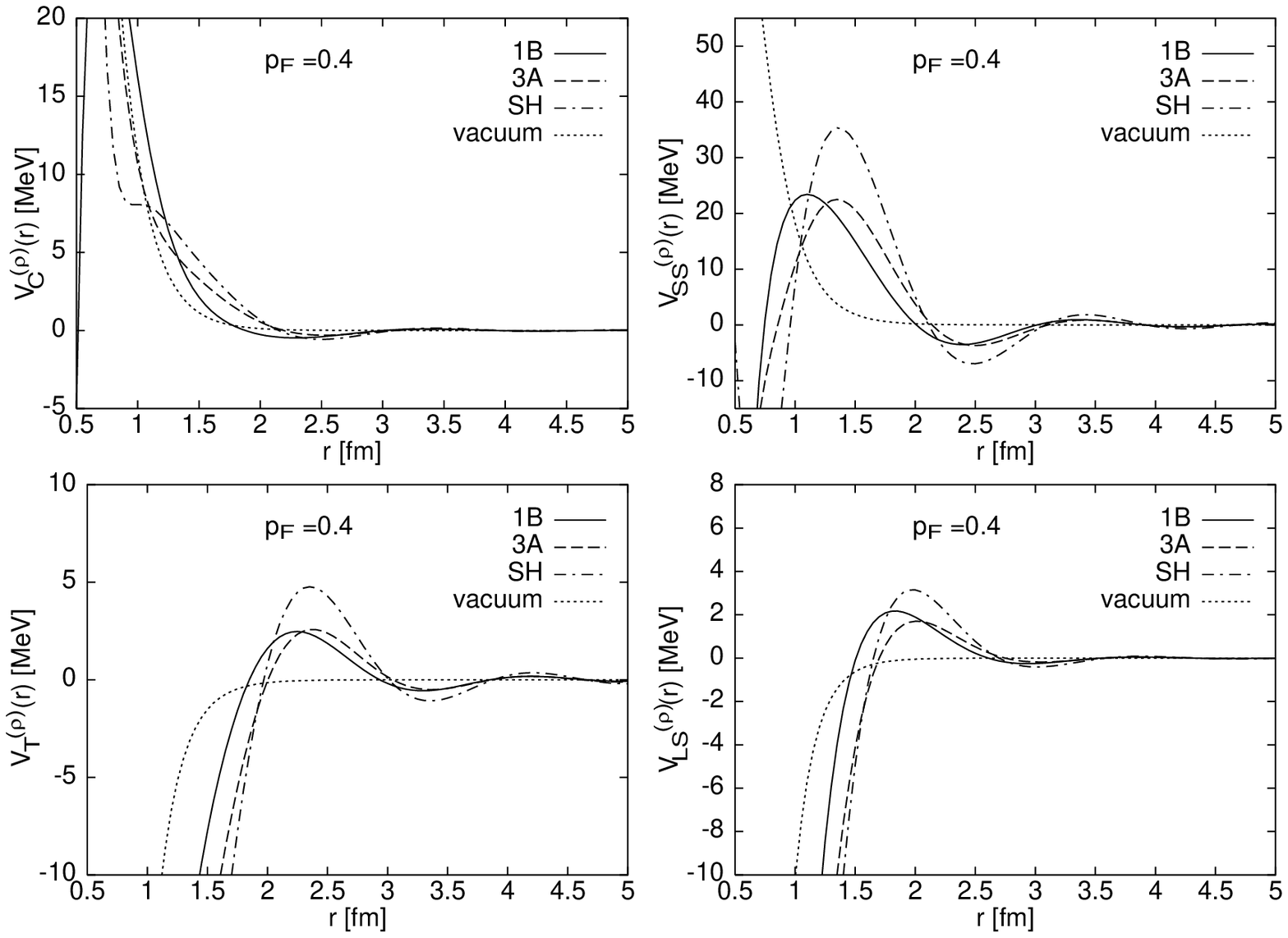}{One-rho exchange component of the potential at
2.4 times saturation density ($p_F$=0.4) for various renormalization schemes:
scheme 1 with parameter set 1B (solid line), scheme 3 with parameter set 
3A (dashed line) and scheme 5 (= ``Shiomi-Hatsuda'') with the parameter 
set of Machleidt Bonn-B potential (dot-dashed line). The potential in vacuum
is also shown for reference (dotted line). The left upper panel
shows the central component, the right upper panel displays the
spin-spin component, the left lower panel corresponds to the tensor
part. Finally the spin-orbit component is represented in the right lower
panel.}{Fig. 9}

\mafigura{16cm}{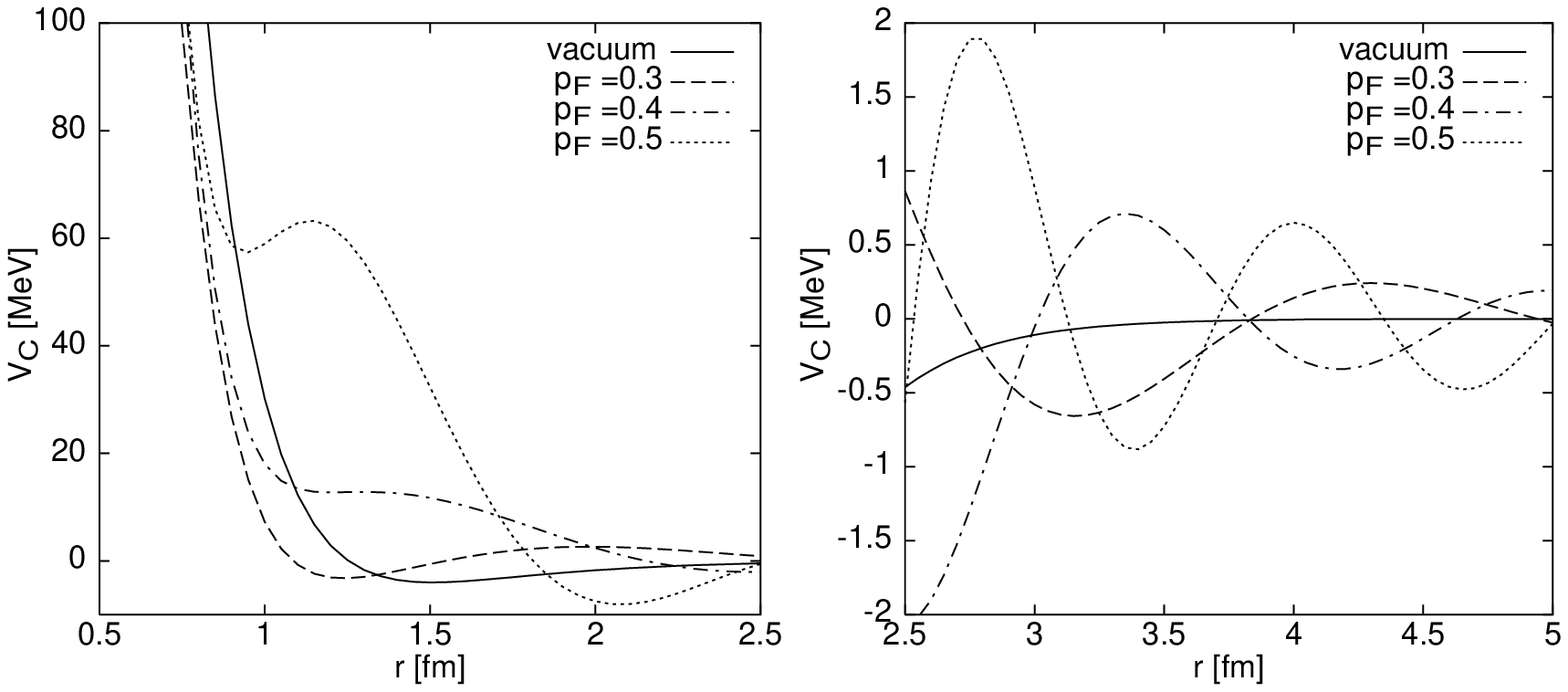}{Central component of the potential including all
($\sigma$, $\omega$, $\rho$, $\pi$) mesons in the medium, using parameter
set 1B with renormalization scheme 1.  We use the same conventions as in the 
left panel to represent various densities. The left panel represents the short 
range part of the potential. The right panel illustrates the oscillatory 
behavior in the long range part.}{Fig. 10}

\mafigura{10cm}{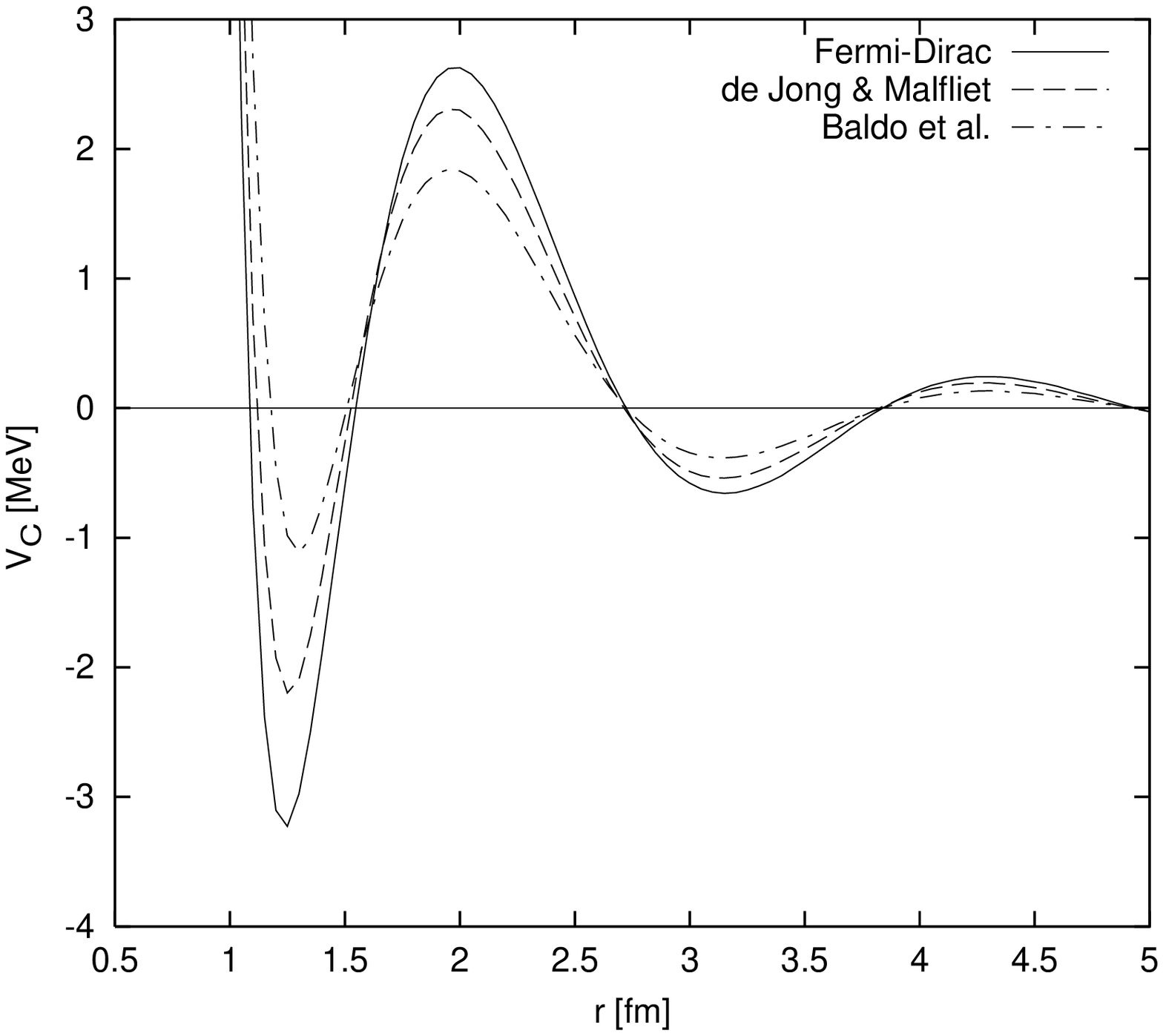}{Influence of the rounding off of the momentum
distribution function on the amplitude of Friedel oscillations.
We show the central component of the potential including all mesons 
at saturation density and vanishing temperature, using parameter set 1B 
with renormalization scheme 1. The solid line was obtained for a Fermi-Dirac 
distribution (as in previous figures). The dashed line uses the parametrization
of the result of a (relativistic) Dirac-Brueckner-Hartree Fock calculation
by de Jong and Malfliet [58]. The dash-dotted line uses the parametrization
of the result of a (nonrelativistic) Brueckner-Hartree Fock calculation
by Baldo {\it et al.} [57].}{Fig. 11}

\mafigura{10cm}{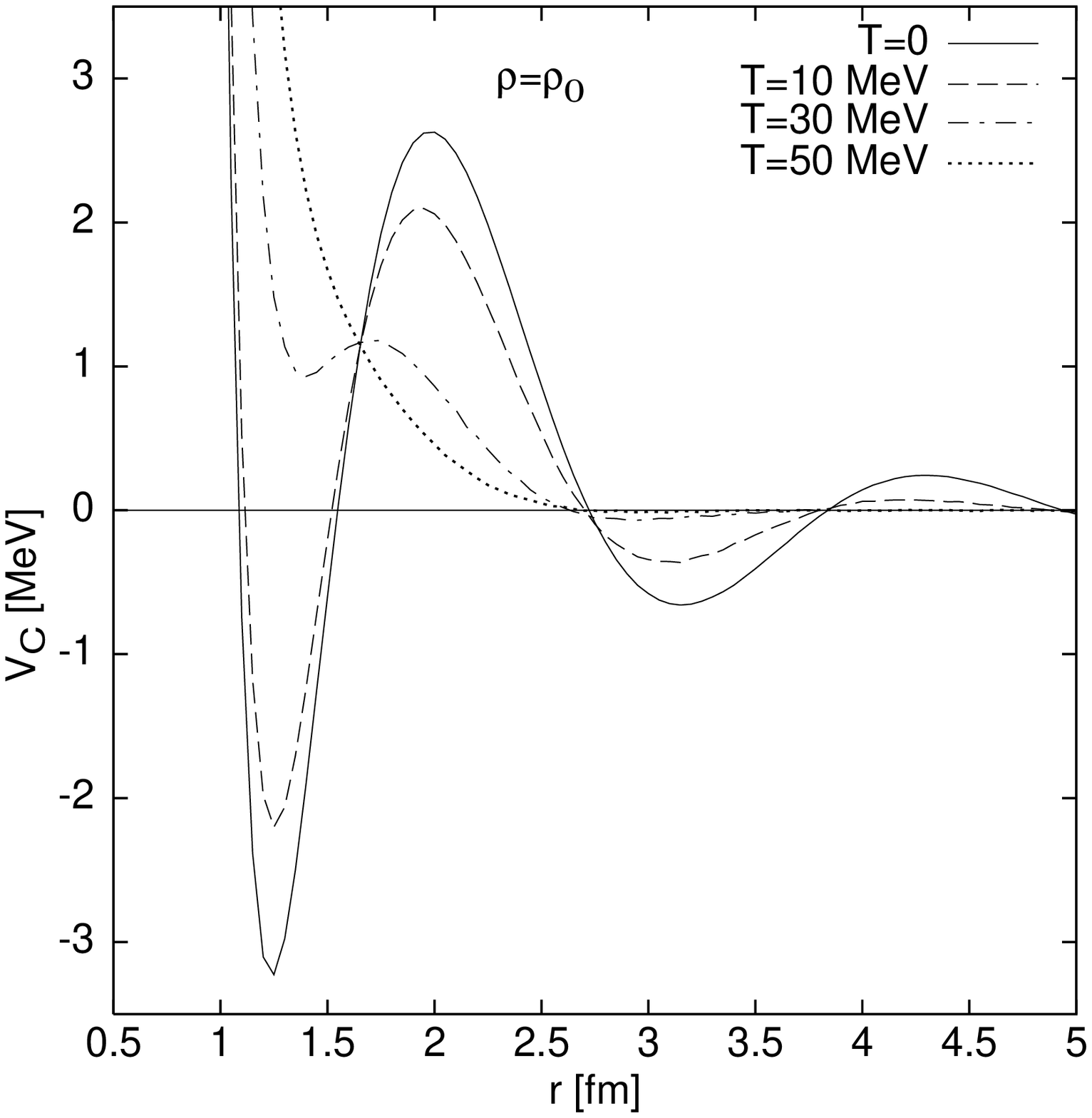}{Influence of temperature on the amplitude of  
oscillations. We show the central component of the potential including all mesons 
at saturation density, using parameter set 1B. The solid, dashed, dot-dashed and
dotted lines correspond to temperatures T=0, 10, 30, 50 MeV respectively.}{Fig. 12}


\begin{thebibliography}{99}

\bibitem{M89}  R. Machleidt. 'The Meson Theory of Nuclear Forces and Nuclear
Structure.' Adv. in Nucl. Phys.{\bf Vol. 19}; J.W. Negele and E. Vogt Edts.
(Plenum Press, New York, 1989.)

\bibitem{[SW86]}  B.D. Serot, J.D. Walecka, 'The Relativistic Nuclear
Many-Body Problem', Adv. in Nucl. Phys.,{\bf Vol. 16}, J.W. Negele and E.
Vogt Edts. ( Plenum Press, New York, 1986).

\bibitem{[Se97]}  B.D. Serot and J.D. Walecka, {\it Int. J. Mod. Phys.} {\bf %
E6} (1997) 515.

\bibitem{[BM90]}  R. Brockmann and R. Machleidt, {\it Phys. Rev.} {\bf C42},
(1990), 1965.

\bibitem{[BM96]}  R. Brockmann and R. Machleidt : 'The Dirac-Brueckner
Approach'. Prepared for 'Open Problems in Nuclear Matter' M. Baldo, ed.
World Singapore. Preprint nucl-th/9612004.

\bibitem{FR52}  J. Friedel, {\it Phil. Mag.} {\bf 43} (1952) 153; \\
J. Friedel, {\it Nuovo Cim.} {\bf 7} (1958) Suppl. 2, 287.

\bibitem{RO60}  T.J. Rowland, {\it Phys. Rev.} {\bf 119} N3 (1960) 900.

\bibitem{KA88}  J. Kapusta and T. Toimela, {\it Phys. Rev.} {\bf D37} (1988)
3731.

\bibitem{FF96}  W. Florkowski and B. Friman, {\it Nucl. Phys.} {\bf A 611}
(1996) 409.

\bibitem{DPS89}  J. Diaz Alonso, A. P\'{e}rez and H. Sivak, {\it Nucl. Phys.}
{\bf A505} (1989) 695.

\bibitem{GDP94}  E. Gallego, J. Diaz Alonso and A. P\'{e}rez, {\it Nucl.
Phys.} {\bf A578} (1994) 542.

\bibitem{GDP94L}  J. Diaz Alonso, E. Gallego and A. P\'erez, {\it Phys. Rev.
Lett.} V.73; N.19; (1994); 2536.

\bibitem{DM98} J. Diaz Alonso and L. Mornas, {\it Nucl. Phys.} {\bf A629} 
(1998) 679.

\bibitem{IP82} N. Iwamoto, C.J. Pethick, {\it Phys. Rev.} {\bf D25} 
(1982) 313.

\bibitem{RPLP99} S. Reddy, M. Prakash, J.M. Lattimer and J.A. Pons,
{\it Phys. Rev.} {\bf C59} (1999) 2888. 

\bibitem{W93} K. Wehrberger, {\it Phys. Rep.} {\bf 225} (1993) 273.

\bibitem{JW94}  A. D. Jackson and T. Wettig, {\it Phys. Rep}. {\bf 273}
(1994) 325.

\bibitem{DS90} J. Dukelsky and P. Schuck, {\it Nucl. Phys.}{\bf A512} (1990) 
466.

\bibitem{BKS99} Th. Bornath, D. Kremp and M. Schlanges, eprint physics/9903039

\bibitem{HCP95}  J.M. H\"auser, W. Cassing and A. Peter, {\it Nucl. Phys.} 
{\bf A585} (1995) 727. \\
J.M. H\"auser, W. Cassing, A. Peter and M.H. Thoma, {\it Z.Phys.} {\bf A353}
 (1996) 301.

\bibitem{CL94} J.C. Caillon and J. Labarsouque, {\it Nucl. Phys.} {\bf A572}
649.

\bibitem{ChR96}  G. Chanfray, R. Rapp and J. Wambach, {\it Phys. Rev. Lett.} 
{\bf 76} (1996) 368.

\bibitem{DP91}  J. Diaz Alonso and A. P\'{e}rez, {\it Nucl. Phys.} 
{\bf A526} (1991) 623.

\bibitem{DA85}  J. Diaz Alonso, {\it Ann. Phys.} {\bf 160, N1} (1985), 1.

\bibitem{CHIN77}  S.A. Chin, {\it Ann. Phys. (N.Y.)} {\bf 108} (1977), 301.

\bibitem{[H78]}  J. Diaz Alonso and R. Hakim, {\it Phys. Lett.} {\bf A 66},
(1978), 476.

\bibitem{[SH94]}  H. Shiomi and T. Hatsuda, {\it Phys. Lett.} {\bf B334}
(1994) 281.

\bibitem{[KS88]} H. Kurasawa, T. Suzuki, {\it Nucl. Phys.} {\bf A490} (1988) 
571.

\bibitem{Sarkar98} S. Sarkar, J. Alam, P. Roy, A. Dutt-Mazumder, 
B. Dutta-Roy and B. Sinha, {\it Nucl. Phys.} {\bf A634} (1998) 206.

\bibitem{DMP00} J. Diaz-Alonso, L. Mornas and M.A. P\'erez-Garc{\'\i}a, 
in preparation

\bibitem{JDA-LM-PLB} J. Diaz Alonso and L. Mornas, {\it Phys. Lett} {\bf B437} 
(1998) 12-18.

\bibitem{LH89} K. Lim and C.J. Horowitz, {\it Nucl. Phys.} {\bf A501} (1989) 729. 

\bibitem{JDAprivate} J. Diaz-Alonso, private communication.

\bibitem{BR91}  G. E. Brown and M. Rho, {\it Phys. Rev. Lett.} {\bf 66}
(1991) 2720.

\bibitem{[HL92]}  T. Hatsuda and Su H. Lee, {\it Phys. Rev.} {\bf C46}
(1992) R34.

\bibitem{Leinweber}  Xuemin Jin and D.B. Leinweber, 
{\it Phys. Rev.} {\bf C52} (1995) 3344.

\bibitem{[TC94]}  D.K. Griegel and T. D. Cohen, {\it Phys. Lett.} {\bf B333}
(1994) 27.

\bibitem{LPM98} S. Leupold, W. Peters and U. Mosel, {\it Nucl.} Phys {\bf A628}
(1008) 

\bibitem{KKW97} F. Klingl, N. Kaiser and W. Weise, {\it Nucl. Phys.} {\bf A624} 
(1997) 527.

\bibitem{Nyff00} A. Nyffeler, LANL eprint hep-ph/0010329  

\bibitem{Oset00} D. Cabrera, E. Oset and M.J. Vicente Vacas, LANL preprint
nucl-th/0011037

\bibitem{RW99} R. Rapp and J. Wambach, LANL preprint hep-ph/9909229 

\bibitem{[ST97]}  K. Saito, K. Tsushima and A.W. Thomas, {\it Phys. Rev.} 
{\bf C55} (1997) 2637; \\
K. Saito, K. Tsushima and A.W. Thomas, {\it Phys. Rev.} {\bf C56} (1997)  566.

\bibitem{[HFN93]} M. Herrmann, B.L. Friman and W. N\"orenberg, {\it Nucl. Phys.} 
{\bf A560} (1993) 411.

\bibitem{klingl} F. Klingl, N. Kaiser and W. Weise, {\it  Z. Phys.} {\bf A356} 
(1996) 193.

\bibitem{[Pi95]}  R.D. Pisarski, {\it Nucl. Phys.} {\bf A590} (1995) 553c; \\
R.D. Pisarski, {\it Phys. Rev.} {\bf D52} (1995) 3773.

\bibitem{[Song]} C. Song, {\it Phys. Rev.} {\bf D48} (1993) 1375

\bibitem{HELIOS} M. Masera {\it el al}, HELIOS collaboration, {\it Nucl. Phys.} 
{\bf A590} (1992) 93c.

\bibitem{CERES} G. Agakichiev {\it et al.}, CERES collaboration, 
{\it Phys. Rev. Lett.} {\bf 75} (1995) 1272; \\
 G. Agakichiev {\it et al.}, CERES collaboration, {\it Phys. Lett.} {\bf B422}
 (1998) 405

\bibitem{LK94}  G.Q. Li and C.M. Ko, {\it Nucl. Phys.} {\bf A582 }(1994) 731.

\bibitem{KB96}  C. M. Ko, G. Q. Li, G. E. Brown and H. Sorge, {\it Nucl.
Phys.} {\bf A610 }(1996) 342c.

\bibitem{[CEK95]}  W. Cassing, W. Ehehalt and C.M. Ko, {\it Phys. Lett.} 
{\bf B363} (1995) 35.

\bibitem{RCW97} R. Rapp, G. Chanfray and J. Wambach, {\it Nucl. Phys.}
{\bf A617} (1997) 472.

\bibitem{[Fri97]}  B. Friman and H.J. Pirner, {\it Nucl. Phys.} {\bf A617}
(1997) 496.

\bibitem{McC92}  J. B. McClelland {\it et al.}, {\it Phys. Rev. Lett.} {\bf 69}
(1992) 582.

\bibitem{BW94}  G. E. Brown and J. Wambach, {\it Nucl. Phys.} {\bf A568}
(1994) 895.

\bibitem{UG94}  W. Unkelbach, C. Glashausser, A. Green and J. Wambach, 
{\it Nucl. Phys.} {\bf A569} (1994) 353c.

\bibitem{Toki} K. Yoshida and H. Toki, {\it Nucl. Phys.} {\bf A648} (1999) 75.

\bibitem{DPS98}  J. Diaz Alonso, A. P\'{e}rez and H. Sivak, hep-ph/9803344.


\bibitem{Pri90}  C. E. Price, J. R. Shepard and J. A. McNeil, {\it Phys.
Rev.} {\bf C 41}, (1990) 1234; \\
C. E. Price, J. R. Shepard and J. A. McNeil, {\it Phys.Rev.} {\bf C 42}, 
(1990) 247.

\bibitem{urban} M. Urban, M. Buballa, R. Rapp, J. Wambach, Nucl. Phys.
{\bf A673} (1999) 357. 

\bibitem{cassingbratk} W. Cassing, E.L. Bratkovskaya, Phys. Rep. 
{\bf 308} (1999) 65.

\bibitem{[Asakawa]} M. Asakawa, C.M. Ko, {\it Phys. Rev.} {\bf C48} (1993) 
R526; \\
M. Asakawa, C.M. Ko, P. Levai, X.J. Qiu, {\it Phys. Rev.} {\bf C46} (1992) 
R1159.

\bibitem{Baldo90} M. Baldo, I. Bombaci, G. Giansiracusa, U. Lombardo, 
C. Mahaux and R. Sartor, {\it Phys. Rev.} {\bf C41} (1990) 1748.   

\bibitem{dJM91} F. de Jong and R. Malfliet, {\it Phys. Rev.} {\bf C44} (1991) 998.  

\bibitem{[DH84]}  J. Diaz Alonso and R. Hakim, {\it Phys. Rev.} {\bf D 29}
(1984), 2690. 

\end{thebibliography}
\end{document}